\documentclass[10pt]{article}
\usepackage{graphicx}
\usepackage{amssymb}
\usepackage{epstopdf}
\usepackage{url}
\newcommand{\compl}{{\mathbb C}}
\newcommand{\real}{{\mathbb R}}

\newcommand{\bibnodot}[1]{}
\title{A Theory of Concepts and Their Combinations II:\vspace{0.3\baselineskip} \\
 \large A Hilbert Space Representation\footnote{To appear in {\it Kybernetes}, Summer 2004.}}
\author{Diederik Aerts\\
        \normalsize\itshape
        Center Leo Apostel for Interdisciplinary Studies \\
         \normalsize\itshape
         Department of Mathematics and Department of Psychology\\
        \normalsize\itshape
        Vrije Universiteit Brussel, 1160 Brussels, 
       Belgium \\
        \normalsize
        E-Mail: \textsf{diraerts@vub.ac.be} \\ \\
		Liane Gabora \\
		 \normalsize\itshape
		 Center Leo Apostel for Interdisciplinary Studies \\
		  \normalsize\itshape
		  Vrije Universiteit Brussel and Department of Psychology\\
		   \normalsize\itshape
		   University of California, Berkeley, CA 94720-1650, USA \\
		   \normalsize
        E-Mail: \textsf{liane@berkeley.edu}}
\date{}
\begin{document}
\maketitle
\begin{abstract}
\noindent
The sets of contexts and properties of a concept are embedded in the complex Hilbert space of quantum mechanics. States are unit vectors or density operators, and contexts and properties are orthogonal projections. The way calculations are done in Hilbert space makes it possible to model how context influences the state of a concept. Moreover, a solution to the combination of concepts is proposed. Using the tensor product, a procedure for describing combined concepts is elaborated, providing a natural solution to the pet fish problem. This procedure allows the modeling of an arbitrary number of combined concepts. By way of example, a model for a simple sentence containing a subject, a predicate and an object, is presented.
\end{abstract} 

\begin{quotation}
\noindent
Keywords: concept, combination, quantum mechanics, Hilbert space, context, entanglement, pet fish problem, tensor product. 
\end{quotation}

%%%%%%%%%%%%%%%%%%%%%%%%%%%%%%%%%%%%%%%%%%%%%%%%%%%%%%%%%%%
%%%%%%%%%%%%%%%%%%%%%%%%%%%%%%%%%%%%%%%%%%%%%%%%%%%%%%%%%%%
\section{Introduction}
The SCOP theory models a concept as an entity that can be in different states such that a state changes under the influence of a context. The notion of `state of a concept' makes it possible to describe a specific contextual effect, namely that an exemplar of the concept has different typicalities and a property of the concept different applicabilities under different contexts. The experiment put forward in \cite{aertsgabora02} illustrates this contextual effect. In this article we present a numerical mathematical model for the representation of a concept, built with a mathematical formalism originally used in quantum mechanics, and we show that the data of the above mentioned experiment can be reproduced by the model. Specifically, the model is built using the Hilbert space of quantum mechanics, states are represented by unit vectors of this Hilbert space and contexts and properties by projection operators, and the change of state under the influence of a context is described by von Neumann's `quantum collapse state transformation' in Hilbert space \cite{vonneumann01}. 

This article deals primarily with the question of what happens when concepts combine. As explained in \cite{aertsgabora02}, known theories of concepts (prototype, exemplar and theory) cannot deliver a model for the description of the combination of concepts. We show that the standard quantum mechanical procedure for the description of the compound of quantum entities, {\it i.e.} the tensor product procedure, delivers a description of how concepts combine. Specifically, given the Hilbert spaces of individual concepts, the combination of these concepts is described by the tensor product Hilbert space of these individual Hilbert spaces, and the quantum formalism applied in this tensor product Hilbert space. In this way we work out an explicit description of the combination of `pet' and `fish' in `pet fish', and show that our model describes the guppy effect, and as a consequence solves in a natural way what has come to be known as the `pet fish problem'.

We were amazed to find that not only combinations of concepts like `pet fish', but also sentences like `the cat eats the food' can be described in our formalism by nonproduct vectors of the tensor product (representing the so called entangled states of quantum mechanics) of the individual Hilbert spaces corresponding to the concepts in the combination. It is quantum entanglement that accounts for the most meaningful combinations of concepts. In the last section of the article we explain the relation between our Hilbert space model of concepts and von Foerster's quantum memory approach.

%%%%%%%%%%%%%%%%%%%%%%%%%%%%%%%%%%%%%%%%%%%%%%%%%%%%%%%%%%%
%%%%%%%%%%%%%%%%%%%%%%%%%%%%%%%%%%%%%%%%%%%%%%%%%%%%%%%%%%%
\section{The Mathematics for a Quantum Model} \label{sec:quantummodel}
This section introduces the mathematical structure necessary to construct a Hilbert space representation of a concept.
 
\subsection{Hilbert Space and Linear Operators} \label{sec:quantummath}
A Hilbert space ${\cal H}$ is a vector space over the set of complex numbers $\compl$, in which case we call it a complex Hilbert space, or the set of real numbers $\real$, in which case we call it a real Hilbert space. Thus the elements of a Hilbert space are vectors. We are interested in finite dimensional complex or real Hilbert spaces and hence do not give a definition of an abstract Hilbert space. Let us denote $\compl^n$ to be the set of $n$-tupels $(x_1, x_2, \ldots, x_{n-1}, x_n)$, where each $x_k$ for $1 \le k \le n$ is a complex number. In a real Hilbert space, the elements $x_k$ are real numbers, and the set of $n$-tupels is denoted $\real^n$. However, we consider the complex Hilbert space case as our default, because the real Hilbert space case is a simplified version of it, and its mathematics follows immediately from the complex case. We define a sum and a multiplication with a complex number as follows:
For $(x_1, x_2, \ldots, x_{n-1}, x_n)$, $(y_1, y_2, \ldots, y_{n-1}, y_n) \in \compl^n$ and $\alpha
\in
\compl$, we have
\begin{eqnarray}
&&(x_1, x_2, \ldots, x_{n-1}, x_n) + (y_1, y_2, \ldots, y_{n-1}, y_n) = (x_1 + y_1, x_2 + y_2, \ldots, x_{n-1} + y_{n-1}, x_n + y_n) \\
&&\alpha (x_1, x_2, \ldots, x_{n-1}, x_n) = (\alpha \cdot x_1, \alpha \cdot x_2, \ldots, \alpha \cdot x_{n-1}, \alpha \cdot x_n)
\end{eqnarray}
This makes $\compl^n$ into a complex vector space. We call the $n$ tupels $(x_1, x_2, 
\ldots, x_{n-1}, x_n)$ vectors and denote them $|x \rangle \in \compl^n$.
We also define an inproduct between vectors of $\compl^n$. For $|x\rangle, |y\rangle 
\in \compl^n$ we have:
\begin{equation}
\langle x|y\rangle = x_1^* \cdot y_1 + x_2^* \cdot y_2 + \ldots + x_{n-1}^* \cdot y_{n-1} + x_n^* \cdot y_n
\end{equation}
where $x_i^*$ is the complex conjugate of $x_i$. The inproduct of two vectors is a complex number,
hence  $\langle x |y\rangle \in \compl$. For $\alpha, \beta \in \compl$ and $|x\rangle, |y\rangle, |z\rangle \in \compl^n$ we have
\begin{equation} 
\langle \alpha x + \beta y|z\rangle = \alpha^*\langle x|z\rangle + \beta^*\langle y|z\rangle \quad {\rm and} \quad  \langle x| \alpha y + \beta z\rangle = \alpha \langle x|y\rangle + \beta \langle y|z\rangle \label{eq:bilinearinproduct}
\end{equation}
This shows that the inproduct is conjugate linear in the first slot, and linear in the second slot of the operation $\langle\ \cdot\ |\ \cdot\
\rangle$. The complex vector space $\compl^n$ equipped with this inproduct is an $n$ dimensional complex Hilbert space. Any $n$ dimensional complex Hilbert space is isomorphic to $\compl^n$.
The inproduct gives rise to a length for vectors and an angle between two vectors, {\it i.e.} for $|x\rangle, |y\rangle \in \compl^n$ we define
\begin{equation} \label{eq:angle}
\|x\| = \sqrt{\langle x|x\rangle} \quad {\rm and} \quad \cos(x,y) = {|\langle x|y\rangle| \over \|x\| \cdot \|y\|}
\end{equation}
Two non-zero vectors $|x\rangle, |y\rangle \in \compl^n$ are said to be orthogonal iff $\langle x|y\rangle = 0$. (\ref{eq:angle})
shows that if the inproduct between two non zero vectors equals zero, the angle between these vectors is 90 degrees. A linear operator $A$ on $\compl^n$ is a function $A: \compl^n \rightarrow \compl^n$ such that
\begin{equation}
A (\alpha |x\rangle + \beta |y\rangle) = \alpha A|x\rangle + \beta A|y\rangle
\end{equation}
For the finite dimensional Hilbert space $\compl^n$ each linear operator $A$ can be fully described by a $n
\times n$ matrix
$A_{ij}$, $1 \le i \le n, 1 \le j \le n$ of complex numbers, where
\begin{equation}
A|x\rangle = (\sum_{j=1}^n A_{1j}x_j, \sum_{j=1}^n A_{2j}x_j, \ldots, \sum_{j=1}^n A_{n-1,j}x_j, \sum_{j=1}^n A_{n,j}x_j)
\end{equation}
if $|x\rangle = (x_1, x_2, \ldots, x_{n-1}, x_n)$. We make no distinction between the linear operator
$A$ and its matrix representation $A_{ij}$. This gives us the necessary ingredients to explain how states, contexts and properties of a concept are represented in the Hilbert space model.

\subsection{States} \label{sec:states}
There are two types of states in quantum mechanics: pure states and density states. A pure states is represented by a unit vector
$|x\rangle \in \compl^n$, {\it i.e.} a vector $|x\rangle \in \compl^n$ such that $\|x\| = 1$. A density state is
represented by a density operator $\rho$ on
$\compl^n$, which is a linear operator 
that is self adjoint. This means that 
\begin{equation}
\rho_{ij} = \rho_{ji}^* \label{eq:selfadjoint}
\end{equation}
for all $i, j$ such that $1 \le i \le n, 1 \le j \le n$. Furthermore, it is semi definite, which means that $\langle x|\rho|x\rangle \ge 0$ $\forall\ |x\rangle \in \compl^n$ and its trace, which is the sum of the diagonal elements of its matrix representation, is equal to 1. Hence $\sum_{i=1}^n\rho_{ii} = 1$. So, to represent the concept `pet' and the situation described previously using this quantum model, we determine the dimension $n$ of the Hilbert space, and represent the states $p_1, p_2, \ldots, p_n \in \Sigma$ of `pet' using unit vectors or density operators of the Hilbert space $\compl^n$.

\subsection{Properties and Weights}
A property in quantum mechanics is represented by means of a linear operator, which is an orthogonal projection operator or an orthogonal projector. An orthogonal projection operator $P$ is also a self adjoint operator; hence (\ref{eq:selfadjoint}) must be satisfied, {\it i.e.},
$P_{ij} = P_{ji}^*$. Furthermore for an orthogonal projector, it is necessary that the square of the operator equals the operator itself. Hence $P^2 = P$. Expressed using the components of the matrix of $P$, this gives $\sum_{j=1}^n P_{ij}P_{jk} = P_{ik}$.

This means that to describe the concept `pet' we need to find two orthogonal projection operators $P_a$
and $P_b$ of the complex Hilbert space $\compl^n$ that represent the properties $a, b \in {\cal L}$.

Let us introduce the quantum mechanical rule for calculating the weights of properties in different states. If the state $p$ is a pure state represented by a unit vector $|x_p\rangle \in \compl^n$ we have
\begin{equation} \label{eq:weight}
\nu(p, a) = \langle x_p|P_a|x_p\rangle 
\end{equation}
If the state $p$ is a density state represented by the density operator $\rho_p$ we have
\begin{equation} \label{eq:quantumweight}
\nu(p, a) = Tr \rho_p P_a
\end{equation}
where $Tr \rho P_a$ is the trace (the sum of the diagonal elements) of the product of operator $\rho$ with operator $P_a$.

\subsection{Contexts, Probabilities and Change of State}
In quantum mechanics, a measurement is described by a linear operator which is a self adjoint operator, hence represented by
an $n \times n$ matrix $M_{ij}$ that satisfies (\ref{eq:selfadjoint}), {\it i.e}, $M_{ij} = M_{ji}^*$. Although it is standard to represent a context---which in the case of physics is generally a measurement---using a self adjoint operator, we will use the set of orthogonal projection operators that form the spectral decomposition of this self adjoint operator, which is an equivalent representation. Note that we have been considering `pieces of context' rather than total contexts, and a pieces of context is represented by one of these projection operators. Hence, a (piece of) context $e$ is represented by a projector $P_e$. Such a context $e$ changes a state $p$ of the concept to state $q$ as follows. If $p$ is a pure state represented by the unit vector $|x_p\rangle \in \compl^n$ we have
\begin{equation} \label{eq:contextchangestate}
|x_q\rangle = {P_e|x_p\rangle \over \sqrt{\langle x_p|P_e|x_p\rangle}}
\end{equation}
where
\begin{equation} \label{eq:contextchangestateprobability}
\mu(q, e, p) = \langle x_p|P_e|x_p\rangle
\end{equation}
is the probability that this change takes place. If $p$ is a density state represented by the density operator $\rho_p$ we have
\begin{equation} \label{eq:quantumchange}
\rho_q = {P_e \rho_p P_e \over Tr \rho_p P_e}
\end{equation}
where
\begin{equation} \label{eq:quantumprobability}
\mu(q, e, p) = Tr \rho_p P_e
\end{equation}
is the probability that this change takes place.

\subsection{Orthonormal Bases and Superpositions} \label{sec:orthonormal}
The representation of a state $p$ by a density operator $\rho_p$ is general enough to include the case of pure states. Indeed, it can be proven that if a density operator is also an orthogonal projector, then it is an orthogonal projector that projects onto one vector.

A set of vectors $B= \{|u\rangle:|u\rangle \in \compl^n\}$ is an orthonormal base of $\compl^n$ iff (1) the set of vectors $B$ is a generating set
for $\compl^n$, which means that each vector of $\compl^n$ can be written as a linear combination, {\it i.e.} superposition, of vectors of
$B$, (2) each of the vectors of $B$ has length equal to 1, {\it i.e} $\langle u |u \rangle = 1$ for each $|u\rangle \in B$, and (3) each two different vectors of $B$ are orthogonal to each other, {\it i.e.} $\langle v| w\rangle = 0$ for $|v\rangle, |w\rangle \in B$ and $|v\rangle \not= |w\rangle$. It can be shown that any orthonormal base of $\compl^n$ contains exactly $n$ elements. Given such an orthonormal base $B$ of $\compl^n$, any vector $|x\rangle \in \compl^n$ can be uniquely written as a linear combination or superposition of the vectors of this base. This means that there exist superposition coefficients $\alpha_u
\in \compl$ such that $|x\rangle = \sum_{|u\rangle \in B} \alpha_u |u\rangle$.
Making use of (\ref{eq:bilinearinproduct}) we have $\langle u|x\rangle = \langle u| \sum_{|v\rangle \in B} \alpha_v |v\rangle = \sum_{|v\rangle \in B} \alpha_v \langle u|v\rangle = \alpha_u$, hence
\begin{equation} \label{eq:diracsuperposition}
|x\rangle = \sum_{|u\rangle \in B} |u\rangle \langle u|x\rangle
\end{equation}
From this it follows that
\begin{equation}
\sum_{|u\rangle \in B} |u\rangle \langle u| = {\bf 1}
\end{equation}
which is called the `resolution of the unity' in Hilbert space mathematics.
Consider the projector that projects on $|u\rangle$ and denote it $P_u$. Suppose that $|x\rangle$
is a unit vector. Then we have $|x\rangle = \sum_{|u\rangle \in B} P_u |x\rangle$. Taking into account (\ref{eq:diracsuperposition}) gives us $P_u = |u\rangle \langle u|$.
We also have $P_u|x\rangle = \alpha_u|u\rangle$
and hence
\begin{equation} \label{eq:prob02}
\langle x|P_u|x\rangle = \alpha_u \alpha_u^* = |\alpha_u|^2
\end{equation}
This proves that the coefficients $\alpha_u$ of the superposition of a unit vector $|x\rangle$ in an orthonormal base
$B$ have a specific meaning. From (\ref{eq:contextchangestateprobability}) and (\ref{eq:prob02}) it follows that they are the
square root of the probability that the state of the concept represented by
$|x\rangle$ changes under the influence of the context represented by $P_u$.

It is easy to see that the quantum model is a specific realization of a SCOP. Consider the complex Hilbert space $\compl^n$, and define $\Sigma_Q = \{\rho_p\ \vert\ \rho_p$ is a density operator of ${\cal H}\}$,
${\cal M}_Q = \{ P_e\ \vert\ P_e$ is an orthogonal projection operator of ${\cal H}\}$,  ${\cal L}_Q = \{ P_a\ \vert\ P_a$ is an orthogonal projection operator of ${\cal H}\}$,
and the functions $\mu$ and $\nu$ such that $\mu_Q(q, e, p) = Tr \rho_p P_e$, $\nu_Q(p, a) = Tr \rho_p P_a$ and $\rho_q = {P_e \rho_p P_e / Tr \rho_p P_e}$,
then $(\Sigma_Q, {\cal M}_Q, {\cal L}_Q, \mu_Q, \nu_Q)$ is a SCOP.

%%%%%%%%%%%%%%%%%%%%%%%%%%%%%%%%%%%%%%%%%%%%%%%%%%%%%%%%%%%
%%%%%%%%%%%%%%%%%%%%%%%%%%%%%%%%%%%%%%%%%%%%%%%%%%%%%%%%%%%
\section{A Hilbert Space Representation of a Concept} \label{sec:quantummodel02}
In this section we explain how the quantum mechanical formalism is used to construct a model for a concept. We limit ourselves to the construction of a model of one concept. In the next section we explain how it is possible to model combinations of two or more concepts. 

\subsection{Basic Contexts and Basic States}
Let us re-analyze the experiment in greater detail, taking into account the structure of SCOP derived in \cite{aertsgabora02}. For this purpose, the states and contexts corresponding to the exemplars considered in
Table 2 of \cite{aertsgabora02} are presented in Table \ref{tab:statescontextexemplars}.\begin{table}[h] 
\begin{center}\small
\begin{tabular}{|l|l|l|l|l|}
\hline
\multicolumn{1}{|c|}{exemplar} &  \multicolumn{1}{|c|}{context} & \multicolumn{1}{|c|}{state} \\
\hline
{\it rabbit} & $e_{13}$ & $p_{13}$\\
\hline
{\it cat}  & $e_{14}$ & $p_{14}$ \\
\hline
{\it mouse}  & $e_{15}$  & $p_{15}$\\ 
\hline
{\it bird}  & $e_{16}$ & $p_{16}$\\
\hline
{\it parrot}  & $e_{17}$ & $p_{17}$ \\
\hline
{\it goldfish}  & $e_{18}$ & $p_{18}$\\
\hline
{\it hamster}  & $e_{19}$ & $p_{19}$\\
\hline
{\it canary}  & $e_{20}$ & $p_{20}$\\
\hline
{\it guppy}  & $e_{21}$ & $p_{21}$\\
\hline
{\it snake}  & $e_{22}$ & $p_{22}$ \\
\hline
{\it spider}  & $e_{23}$ & $p_{23}$\\
\hline
{\it dog}  & $e_{24}$ & $p_{24}$ \\
\hline
{\it hedgehog}  & $e_{25}$ &$p_{25}$\\
\hline
{\it guinea pig}  & $e_{26}$ & $p_{26}$ \\
\hline
\end{tabular}
\end{center}
\caption{States and contexts relevant to exemplars of the concept `pet'.} \label{tab:statescontextexemplars}
\end{table}
So, for example, $e_{19}$ is the context `The pet is a hamster', and $p_{15}$ is the state of `pet' under the context $e_{15}$, `The pet is a mouse'. 
In the experiment, subjects were asked to estimate the frequency of a specific exemplar of `pet' given a specific context; for example, the exemplar {\it cat} for the context $e_1$, `The pet is chewing a bone', the frequency of the exemplar {\it dog} for the context $e_2$, `The pet is being taught', {\it etc \ldots}. These estimates guide how we embed the SCOP into a Hilbert space. The hypothesis followed in the construction of the embedding is that the frequency estimates reflect the presence of contexts that are stronger than those explicitly considered in the model, and the distribution of these contexts reflects the frequencies measured in the experiment. Let us call these contexts {\it basic contexts}.
For example the contexts 
\begin{eqnarray}
e_{27},&&{\rm `I\ remember\ how\ I\ have\ seen\ my\ sister\ trying\ to\ teach} \nonumber\\
&&{\rm her\ dog\ to\ jump\ over\ the\ fence\ on\ command'} \\
e_{28},&&{\rm `A\ snake\ as\ pet,\ oh\ yes,\ I\ remember\ having\ seen\ that\ weird\ guy} \nonumber\\ 
&&{\rm on\ television\ with\ snakes\ crawling\ all\ over\ his\ body'} \\
e_{29}, &&{\rm `That\ is\ so\ funny,\ my\ friend\ is\ teaching\ his\ parrot\ to\ say\ my\ name} \nonumber \\ 
&&{\rm when\ I\ come\ in'}
\end{eqnarray}
could be such basic
contexts. And indeed we have $e_{27} \le e_2$ and $e_{27} \le e_{24}$, $e_{28} \le e_4$ and $e_{28} \le e_{22}$, and, $e_{29} \le e_5$ and $e_{29} \le e_{17}$, which shows that these contexts are stronger than any of those considered in the model.
Let us denote $X$ the set of such basic contexts for the concept `pet'.

Here we see how our model integrates similarity based and theory based approaches. The introduction of this set of contexts might give the impression that basic contexts play somewhat the same role as exemplars play in exemplar models. This is however not the case; we do not make claims about whether basic contexts are stored in memory. It is possible, for example, that it is a mini-theory that is stored in memory, a mini-theory that has grown out of the experience a subject has had with (part of) the basic contexts, and hence incorporates knowledge about aspects (for example frequency of appearance in different contexts) of the basic contexts in this way. But it is also possible that some basic contexts are stored in memory. At any rate, they play a structural role in our model, a role related directly to the concept itself. To clarify this, compare their status to the status of a property. The property $a_7$, {\it can swim} is a property of the concept `goldfish' independent of the choice of a specific theory of concept representation, or independent of what is or is not stored in memory. 

We now introduce some additional hypotheses. First, we suppose that each basic context is an atomic context of ${\cal M}$. This means that we stop refining the model with basic contexts; it amounts to demanding that there are no stronger contexts available in the model. They are the most concrete contexts we work with. As mentioned in Section 3.5 of \cite{aertsgabora02}, even if a context is an atomic context, there still might be several eigenstates of this context. As an additional hypothesis, we demand that each basic context has only one eigenstate in the model. This means that also on the level of states we want the basic contexts to describe the most refined situation. Indeed, if an atomic context has different eigenstates, the states penetrate more deeply into the refinement of the model than the contexts do. So our demand reflects an equilibrium in fine structure between states and contexts. The set of eigenstates of the atomic contexts we denote $U$, and we call the elements of $U$ {\it basic states}. The basic states and contexts are not necessarily possible instances of the concept, but an instance can play the role of a basic state and context. Basic states and contexts can be states and contexts that the subject has been confronted with in texts, movies, dreams, conversations, {\it etc \ldots}. Let us introduce
\begin{equation}
E_i = \{u\ \vert\ u \le e_i, u \in X\} \quad {\rm and} \quad X_{ij} = \{u\ \vert\ u \le e_i \wedge e_j, u \in X\}
\end{equation}
where $E_i$ is the set of basic contexts that is stronger or equal to
$e_i$, and $X_{ij}$ the set of basic contexts stronger or equal to $e_i \wedge e_j$. It is easy to prove that $X_{ij} = E_i \cap E_j$.
Indeed, we have $u \in X_{ij} \Leftrightarrow u \le e_i \wedge e_j \Leftrightarrow u \le e_i$ and $u \le e_j \Leftrightarrow u \in E_i \cap E_j$.
Suppose that $n$ is the total number of basic contexts. Let us denote $n(X_{ij})$ the number of basic contexts contained in
$X_{ij}$ and 
$n(E_i)$ the number of basic contexts contained in
$E_i$.
We choose $n(X_{ij})$ and $n(E_i)$ as in Table \ref{tab:basiccontexts} (we have denoted $n(X_{ij})$ as $n_{ij}$ in Table
\ref{tab:basiccontexts}).

\begin{table}
\begin{center}
\small
\begin{tabular}{|l||l|l|l|l|l|l|l|l|l|l|l|l|l|l|} \hline 
\multicolumn{1}{|c||}{exemplar} &  \multicolumn{1}{|c|}{$e_1$} & \multicolumn{1}{|c|}{$e_2$} &
\multicolumn{1}{|c|}{$e_3$} &
\multicolumn{1}{|c|}{$e_4$} & \multicolumn{1}{|c|}{$e_5$} & \multicolumn{1}{|c|}{$e_6$} & \multicolumn{1}{|c|}{$1$}\\
\hline   
& $n(E_1) = 303$  & $n(E_2) = 495$ & $n(E_3) = 500$ & $n(E_4) = 101$ & $n(E_5) = 200$ & $n(E_6) = 100$ & n = 1400  \\ \hline 
{\it rabbit}  & $n_{13,1} = 12$   & $n_{13,2} = 35$   & $n_{13,3} = 75$ & $n_{13,4} = 5$  & $n_{13,5} = 2$ &$n_{13,6} = 0$ & $n(E_{13}) = 98$ \\ 
\hline 
{\it cat}  & $n_{14,1} = 75$  & $n_{14,2} = 65$    & $n_{14,3} = 110$  & $n_{14,4} = 3$  & $n_{14,5} = 6$  & $n_{14,6} = 1$ & $n(E_{14}) = 168$\\ 
\hline 
{\it mouse}   & $n_{15,1} = 9$   & $n_{15,2} = 30$   & $n_{15,3} = 40$ & $n_{15,4} = 11$ & $n_{15,5} = 2$ & $n_{15,6} = 0$ & $n(E_{15}) = 70$\\ 
\hline 
{\it bird}  & $n_{16,1} = 6$   & $n_{16,2} = 40$   & $n_{16,3} = 10$ & $n_{16,4} = 4$ & $n_{16,5} = 34$ & $n_{16,6} = 1$ & $n(E_{16}) = 112$\\ 
\hline 
{\it parrot}  & $n_{17,1} = 6$   & $n_{17,2} = 80$   & $n_{17,3} = 5$  & $n_{17,4} = 4$ & $n_{17,5} = 126$ & $n_{17,6} = 1$ & $n(E_{17}) = 98$\\ 
\hline 
{\it goldfish}  & $n_{18,1} = 3$    &  $n_{18,2} = 10$     & $n_{18,3} = 0$ & $n_{18,4} = 2$ & $n_{18,5} = 0$ & $n_{18,6} = 48$ & $n(E_{18}) = 140$ \\
\hline 
{\it hamster}  & $n_{19,1} = 12$    & $n_{19,2} = 35$     & $n_{19,3} = 30$ & $n_{19,4} = 4$ & $n_{19,5} = 2$ & $n_{19,6} = 0$ & $n(E_{19}) = 98$ \\ 
\hline
{\it canary}   & $n_{20,1} = 3$   & $n_{20,2} = 35$   & $n_{20,3} = 5$  & $n_{20,4} = 2$ & $n_{20,5} = 14$ & $n_{20,6} = 1$ & $n(E_{20}) = 112$ \\
\hline 
{\it guppy}    & $n_{21,1} = 3$    & $n_{21,2} = 10$    & $n_{21,3} = 0$ & $n_{21,4} = 2$  & $n_{21,5} = 0$ & $n_{21,6} = 46$ & $n(E_{21}) = 126$ \\
\hline 
{\it snake}  & $n_{22,1} = 6$    & $n_{22,2} = 10$   & $n_{22,3} = 5$  &  $n_{22,4} = 22$ & $n_{22,5} = 0$ & $n_{22,6} = 1$ & $n(E_{22}) = 42$ \\ 
\hline
{\it spider}   & $n_{23,1} = 3$   & $n_{23,2} = 5$     & $n_{23,3} = 15$ & $n_{23,4} = 23$ & $n_{23,5} = 0$ & $n_{23,6} = 0$ & $n(E_{23}) = 28$ \\
\hline 
{\it dog}  & $n_{24,1} = 150$    & $n_{24,2} = 95$    & $n_{24,3} = 120$ & $n_{24,4} = 3$ & $n_{24,5} = 12$ & $n_{24,6} = 1$ & $n(E_{24}) = 168$ \\
\hline  
{\it hedgehog}   & $n_{25,1} = 6$  & $n_{25,2} = 10$     & $n_{25,3} = 40$  & $n_{25,4} = 12$ & $n_{25,5} = 0$ & $n_{25,6} = 0$ & $n(E_{25}) = 42$\\
\hline  
{\it guinea pig}  & $n_{26,1} = 9$    & $n_{26,2} = 35$   & $n_{26,3} = 45$ & $n_{26,4} = 4$ & $n_{26,5} = 2$ & $n_{26,6} = 0$ & $n(E_{26}) = 98$ \\
\hline
\end{tabular}
\end{center}
\caption{Choice of the distribution of the different types of basic contexts for the concept `pet'.} \label{tab:basiccontexts}
\end{table}

\subsection{Embedding in the Hilbert Space}
We consider a Hilbert space of dimension 1400, hence $\compl^n$, with $n = 1400$. Each basic context $u \in X$ is represented by a projector $|u\rangle \langle u|$, where $|u\rangle \in \compl^n$ is a unit vector, and such that $B = \{|u\rangle\
\vert u
\in X\}$ is an orthonormal base of the Hilbert space $\compl^n$, and the corresponding basic state $u \in U$ is represented by this unit vector
$|u\rangle
\in B$. The ground state
$\hat{p}$ of the concept `pet' is represented by a unit vector
$|x_{\hat{p}}\rangle$, superposition of the base states $B = \{|u\rangle\ \vert\ u \in X\}$
\begin{equation}
|x_{\hat{p}}\rangle = \sum_{u \in X} \alpha_u |u\rangle \quad {\rm where} \quad \alpha_u = \langle u|x_{\hat{p}}\rangle
\end{equation}
$|\alpha_u|^2$ is the probability that the concept `pet' changes to be in base state $|u\rangle$ under context $u$. We write
\begin{equation}
|\alpha_u|^2 = {1 \over 1400} \quad \forall \ u \in X
\end{equation}
This means that each of the basic states $u \in U$
is considered to have an equal probability of being elicited.
We can rewrite the ground state
$\hat{p}$ of `pet' more explicitly now
\begin{equation} \label{eq:stateforpet}
|x_{\hat{p}}\rangle = \sum_{u \in X} {1 \over \sqrt{1400}} |u\rangle
\end{equation}
This means that if the concept `pet' is in its ground state $\hat{p}$, there is a probability of 1/1400 that one of the contexts $u \in X$ acts as a basic context of `pet', and changes the ground state of `pet' to the basic state $u \in U$ of `pet'. This means that for `pet' in its ground state, the probability that a basic context that is contained in $E_i$ gets activated and changes the ground state of `pet' to the corresponding basic state, is given by $n(E_i)/1400$, where $n(E_i)$ is given in Table \ref{tab:basiccontexts}. Let us show that a straightforward calculation proves that this gives exactly the weights in Table 2 of \cite{aertsgabora02}.
Following Table \ref{tab:basiccontexts}, in 98 of the 1400 basic contexts  the pet is a hamster. This means that the weight of {\it hamster} in the ground state of `pet' is 98/1400 = 0.07, which indeed corresponds with what we find in Table 2 of \cite{aertsgabora02} for {\it hamster}. In 28 of the 1400 basic contexts, the pet is a {\it spider}. Hence the weight of {\it spider} in the ground state of `pet' is 28/1400 = 0.02, as in Table 2 of \cite{aertsgabora02}. There are 168 of the 1400 basic contexts where the pet is a {\it dog}, which means that the weight for {\it dog} is 168/1400 = 0.12, as in Table 2 of \cite{aertsgabora02}.

Now that we have introduced the mathematical apparatus of the quantum model, we can show explicitly how a context changes the state of the concept to another state, and the model remains predicting the data of the experiment. Consider the concept `pet' and the context $e_1$,
`The pet is chewing a bone'. The context
$e_1$ is represented by the projection operator $P_{e_1}$ given by
\begin{equation}
P_{e_1} = \sum_{u \in E_1} |u\rangle \langle u|
\end{equation}
where $E_1$ is the set of basic contexts that is stronger than or equal to $e_1$, hence
$E_1 = \{u\ \vert\ u \le e_1, u \in X\}$. Let us calculate the new state $|x_{p_1}\rangle$ that $|x_{\hat{p}}\rangle$ changes to under the influence of $e_1$.
Following (\ref{eq:contextchangestate}) we have
\begin{equation}
|x_{p_1}\rangle = {P_{e_1} |x_{\hat{p}}\rangle \over \sqrt{\langle x_{\hat{p}}| P_{e_1} |x_{\hat{p}}\rangle}}
\end{equation}
Let us calculate this new state explicitly. We have
\begin{equation}
P_{e_1} |x_{\hat{p}}\rangle = \sum_{u \in E_1} |u\rangle \langle u|x_{\hat{p}}\rangle = \sum_{u \in E_1} {1 \over \sqrt{1400}}|u\rangle
\end{equation}
and
\begin{equation}
\langle x_{\hat{p}}| P_{e_1} |x_{\hat{p}}\rangle = \sum_{u \in E_1} \langle x_{\hat{p}}|u\rangle \langle u| x_{\hat{p}}\rangle = \sum_{u \in
E_1} |\langle x_{\hat{p}}|u\rangle|^2 = \sum_{u \in E_1}{1 \over 1400} = {303 \over 1400}
\end{equation}
This gives
\begin{equation}
|x_{p_1}\rangle = \sum_{u \in E_1} {1 \over \sqrt{303}} |u\rangle
\end{equation}

\subsection{Different States and Different Weights} \label{sec:states/weights}
We can now show how the quantum model predicts different weights for the contexts corresponding to the different exemplars in the experiment. Consider for example the context $e_{14}$, `The pet is a cat', and the corresponding state $p_{14}$, `The pet is a cat', and calculate the probability that $p_1$ collapses to $p_{14}$ under context $e_{14}$. First we must calculate the orthogonal projection operator of the Hilbert space that describes $e_{14}$. This projection operator is given by
\begin{equation}
P_{e_{14}} = \sum_{u \in E_{14}} |u\rangle \langle u|
\end{equation}
where $E_{14} = \{u\ \vert\ u \le e_{14}, u \in X\}$. Following the quantum mechanical calculation in
(\ref{eq:contextchangestateprobability}), we get the weight of the exemplar {\it cat} under context $e_1$, {\it i.e.} the probability that state
$p_1$ collapses to state $p_{14}$ under context $e_{14}$, `The pet is a cat'. We have
\begin{equation}
\mu(p_{14}, e_{14}, p_1) = \langle x_{p_1}|P_{e_{14}}|x_{p_1}\rangle
\end{equation}
which gives
\begin{eqnarray}
\langle x_{p_1}|P_{e_{14}}|x_{p_1}\rangle &=& \sum_{u \in E_{14}} \langle x_{p_1} |u\rangle \langle u| x_{p_1}\rangle = \sum_{u \in E_{14}} \sum_{v \in E_1}
\sum_{w \in E_1} {1 \over 303} \langle v|u\rangle \langle u|w\rangle \\
&=& \sum_{u \in E_{14}} \sum_{v \in E_1}
\sum_{w \in E_1} {1 \over 303} \delta(v,u)\delta(u,w) = \sum_{u \in E_1 \cap E_{14}} {1 \over 303} =
{75 \over 303} = 0.25
\end{eqnarray}
corresponding with the experimental result in Table 2 of \cite{aertsgabora02}.
In contrast, let us calculate the weight of the exemplar {\it cat} for `pet' in the ground state $\hat{p}$. Applying the same formula (\ref{eq:contextchangestateprobability}) we have
\begin{equation}
\mu(p_{14}, e_{14}, \hat{p}) = \langle x_{\hat{p}}|P_{e_{14}}|x_{\hat{p}}\rangle
\end{equation}
and

\begin{equation}
\langle x_{\hat{p}}|P_{e_{14}}|x_{\hat{p}}\rangle = \sum_{u \in E_{14}} \langle x_{\hat{p}} |u\rangle \langle u| x_{\hat{p}}\rangle = \sum_{u \in E_{14}}
 {1 \over 1400} = {168 \over 1400} = 0.12
\end{equation}
This also corresponds to the experimental results in Table 2 of \cite{aertsgabora02}.

Let us make some more calculations of states and weights corresponding to exemplars and contexts of the experiment. Consider the context $e_6$, `The pet is a fish'. This context $e_6$ is represented by the projection operator $P_{e_6}$ given by
\begin{equation}
P_{e_6} = \sum_{u \in E_6} |u\rangle \langle u|
\end{equation}
where $E_6$ is the set of basic contexts that is stronger than or equal to $e_6$. Hence $E_6 = \{u\ \vert\ u \le e_6, u \in X\}$. Following (\ref{eq:contextchangestate}) we get the following expression for the state $|x_{p_6}\rangle$
\begin{equation}
|x_{p_6}\rangle = {P_{e_6} |x_{\hat{p}}\rangle \over \sqrt{\langle x_{\hat{p}}| P_{e_6} |x_{\hat{p}}\rangle}}
\end{equation}
We have
\begin{equation}
P_{e_6} |x_{\hat{p}}\rangle = \sum_{u \in E_6} |u\rangle \langle u|x_{\hat{p}}\rangle = \sum_{u \in E_6} {1 \over \sqrt{1400}}|u\rangle
\end{equation}
and
\begin{equation}
\langle x_{\hat{p}}| P_{e_6} |x_{\hat{p}}\rangle = \sum_{u \in E_6} \langle x_{\hat{p}}|u\rangle \langle u| x_{\hat{p}}\rangle = \sum_{u \in
E_6} |\langle x_{\hat{p}}|u\rangle|^2 = \sum_{u \in E_6}{1 \over 1400} = {100 \over 1400}
\end{equation}
This gives
\begin{equation} \label{eq:statepetisfish}
|x_{p_6}\rangle = \sum_{u \in E_6} {1 \over \sqrt{100}} |u\rangle
\end{equation}
Suppose we want to calculate the weights of the exemplar `hedgehog' for this state.
Again using formula (\ref{eq:contextchangestateprobability}) we get
\begin{equation}
\mu(p_{25}, e_{25}, p_6) = \langle x_{p_6}|P_{e_{25}}|x_{p_6}\rangle
\end{equation}
From Table \ref{tab:basiccontexts} follows that $n_{25,6} = 0$, which means that $E_{25} \cap E_6 = \emptyset$. We have no basic contexts in
our model where the pet is a fish and a hedgehog. This means that $P_{e_{25}} \perp |x_{p_6}\rangle$, and hence $P_{e_{25}}|x_{p_6}\rangle = |0\rangle$. As
a consequence we have $\mu(p_{25}, e_{25}, p_6) = 0$, which corresponds to the experimental result in Table 2 of \cite{aertsgabora02}.

Let us calculate the weight for the exemplar {\it goldfish} in the state $p_6$. We have
\begin{equation}
\mu(p_{18}, e_{18}, p_6) = \langle x_{p_6}|P_{e_{18}}|x_{p_6}\rangle
\end{equation}
where
\begin{equation}
P_{e_{18}} = \sum_{u \in E_{18}} |u\rangle \langle u|
\end{equation}
and $E_{18} = \{u\ \vert\ u \le e_{18}, u \in X\}$. Following
(\ref{eq:contextchangestateprobability}) this gives
\begin{eqnarray} \label{eq:weightgoldfishpetfish}
\langle x_{p_6}|P_{e_{18}}|x_{p_6}\rangle &=& \sum_{u \in E_{18}} \langle x_{p_6} |u\rangle \langle u| x_{p_6}\rangle = \sum_{u \in E_{18}} \sum_{v \in E_6}
\sum_{w \in E_6} {1 \over 100} \langle v|u\rangle \langle u|w\rangle \\
&=& \sum_{u \in E_{18}} \sum_{v \in E_6}
\sum_{w \in E_6} {1 \over 100} \delta(v,u)\delta(u,w) = \sum_{u \in E_{18} \cap E_6} {1 \over 100} =
{48 \over 100} = 0.48
\end{eqnarray}
corresponding with the experimental result in Table 2 of \cite{aertsgabora02}.

The foregoing calculations show that our SCOP theory in Hilbert space is able to model the experimental data of the experiment put forward in Section 2.2 of \cite{aertsgabora02}. The choice of the distribution of the basic contexts and states as presented in Table \ref{tab:basiccontexts}, and the corresponding dimension of the Hilbert space, is crucial for the model to predict that experimental data of the experiment. It is possible to see that the distribution of basic contexts and states (Table \ref{tab:basiccontexts}) corresponds more or less to a set theoretical model of the experimental data, such that the Hilbert space model can considered to be a quantization, in the sense used in quantum mechanics, of this set theoretical model.

%%%%%%%%%%%%%%%%%%%%%%%%%%%%%%%%%%%%%%%%%%%%%%%%%%%%%%%%%%%
%%%%%%%%%%%%%%%%%%%%%%%%%%%%%%%%%%%%%%%%%%%%%%%%%%%%%%%%%%%
\section{Combinations of Concepts in the SCOP Model} \label{sec:conceptualcombination}
The previous section explained how to build a model of one concept. This section shows that conceptual combinations can be described naturally using the tensor product of the corresponding Hilbert spaces, the procedure to describe compound entities in quantum mechanics. We give an explicit model for the combinations of the concepts `pet' and `fish', and show how the pet fish problem is thereby solved. Then we illustrate how combinations of more than two concepts can
be described. First we need to explain what the tensor product is.

\subsection{The Tensor Product and Entanglement} \label{sec:tensorproduct}
Consider two quantum entities $S$ and $T$ described respectively in Hilbert spaces ${\cal H}^S$ and ${\cal H}^T$. In quantum mechanics there exists a well known procedure to describe the compound $S \otimes T$ of two quantum entities $S$ and $T$ by means of the Hilbert space ${\cal H}^S \otimes {\cal H}^T$, which is the tensor product of the Hilbert spaces ${\cal H}^S$ and ${\cal H}^T$. The tensor product behaves like a product; for example, take $\alpha \in \compl$, $|x^S\rangle \in {\cal H}^S$ and $|x^T\rangle \in
{\cal H}^T$, then we have
\begin{equation}
\alpha (|x^S\rangle \otimes |x^T\rangle) = (\alpha |x^S\rangle) \otimes |x^T\rangle = |x^S\rangle \otimes (\alpha |x^T\rangle)
\end{equation}
However, it is not commutative, meaning that even when a Hilbert space is tensored with itself, for $|x\rangle \in {\cal H}$ and $|y\rangle \in
{\cal H}$ we have
$|x\rangle
\otimes |y\rangle \in {\cal H} \otimes {\cal H}$ is in general not equal to $|y\rangle \otimes |x\rangle$. The mathematical construction of the tensor product in all its details is not trivial. The best way to imagine what the tensor product space is like is to consider two orthonormal bases $B^S$ and $B^T$ respectively of the subspaces ${\cal H}^S$ and ${\cal H}^T$ and note that the set of vectors $\{|u^S\rangle \otimes |u^T\rangle: |u^S\rangle \in B^S, |u^T\rangle \in B^T \}$ is an
orthonormal base of the tensor product
${\cal H}^S \otimes {\cal H}^T$. Concretely this means that each vector $|z\rangle \in {\cal H}^S \otimes {\cal H}^T$ can be written as a linear combination of elements of this orthonormal base
\begin{equation} \label{eq:vectortensorproduct}
|z\rangle = \sum_{|u^S\rangle \in B^S, |u^T\rangle \in B^T} \alpha_{u^S,u^T} |u^S\rangle \otimes |u^T\rangle
\end{equation}
We need to
explain some of the more sophisticated aspects of the tensor product, because they are crucial for the description of conceptual combinations. The first aspect is that vectors of the tensor product can be product vectors or nonproduct vectors. The difference between them can be illustrated with a simple example. Consider the tensor product $\compl^2 \otimes
\compl^2$, and two vectors
$|x\rangle, |y\rangle
\in
\compl^2$, and their tensor product $|x\rangle \otimes |y\rangle \in \compl^2 \otimes \compl^2$. Suppose further that $|u\rangle_1, |u\rangle_2$ is
an orthonormal base of $\compl^2$, which means that we can write
\begin{equation}
|x\rangle = \alpha |u\rangle_1 + \beta |u\rangle_2 \quad {\rm and} \quad |y\rangle = \gamma |u\rangle_1 + \delta|u\rangle_2
\end{equation}
which gives
\begin{eqnarray}
|x\rangle \otimes |y\rangle &=& (\alpha |u\rangle_1 + \beta |u\rangle_2) \otimes (\gamma |u\rangle_1 + \delta |u\rangle_2) \\
&=& \alpha \gamma |u\rangle_1 \otimes |u\rangle_1 + \alpha \delta |u\rangle_1 \otimes |u\rangle_2 + \beta \gamma |u\rangle_2 \otimes |u\rangle_1 +
\beta \delta |u\rangle_2 |u\rangle_2
\end{eqnarray}
Taking into account the uniqueness of the decomposition in (\ref{eq:vectortensorproduct}) we have
\begin{equation}
|x\rangle \otimes |y\rangle = \alpha_{11} |u\rangle_1 \otimes |u\rangle_1 + \alpha_{12} |u\rangle_1 \otimes |u\rangle_2 + \alpha_{21}
|u\rangle_2
\otimes |u\rangle_1 +
\alpha_{22} |u\rangle_2 \otimes |u\rangle_2
\end{equation} 
with
\begin{equation} \label{eq:productrequirements}
\alpha_{11} = \alpha \gamma \quad \alpha_{12} = \alpha \delta \quad \alpha_{21} = \beta \gamma \quad \alpha_{22} = \beta \delta
\end{equation}
It is easy to see that an arbitrary vector $|z\rangle \in \compl^2 \otimes \compl^2$ is not always of the form $|x\rangle \otimes |y\rangle$. For example, choose 
\begin{equation}
|z\rangle = |u\rangle_1 \otimes |u\rangle_1 + |u\rangle_2 \otimes |u\rangle_2
\end{equation}
This amounts to choosing in the decomposition of $|z\rangle$ following formula (\ref{eq:vectortensorproduct}) $\alpha_{11} = \alpha_{22} = 1$ and
$\alpha_{12} = \alpha_{21} = 0$. If $|z\rangle$ chosen in this way were equal to a product vector like $|x\rangle \otimes |y\rangle$, we would
find $\alpha, \beta, \gamma, \delta \in \compl$ such that (\ref{eq:productrequirements}) are satisfied. This means that
\begin{equation} \label{eq:nonproductstate}
\alpha \gamma = \beta \delta = 1 \quad {\rm and} \quad \alpha \delta = \beta \gamma = 0
\end{equation}
This is not possible; there does not exist $\alpha, \beta, \gamma, \delta$ that satisfy (\ref{eq:nonproductstate}). Indeed, suppose that $\alpha
\delta = 0$, then one of the two $\alpha$ or $\delta$ has to equal zero. But then one of the two $\alpha \gamma$ or $\beta \delta$ also equals zero,
which proves that they cannot both equal 1, as demanded in (\ref{eq:nonproductstate}). This proves that $|z\rangle = |u\rangle_1 \otimes |u\rangle_1
+ |u\rangle_2 \otimes |u\rangle_2$ is a nonproduct vector, {\it i.e.} it cannot be written as the product of a vector in $\compl^2$ with another
vector in $\compl^2$. 

Nonproduct vectors of the tensor product Hilbert space represent nonproduct states of the compound concept described by this tensor product Hilbert space. It is these nonproduct states that contain entanglement, meaning that the effect of a context on one of the two sub-entities (sub-concepts) also influences the other sub-entity (sub-concept) in a specific way. As we will see, it is also these nonproduct states that make it possible to represent the relation of entanglement amongst sub-concepts
as one of ways concepts can combine. Specifically (as we will show explicitly in Section \ref{sec:petfishentangled}) combinations like `pet fish' are described as entangled (nonproduct) states of `pet' and `fish' within the tensor product of their respective Hilbert spaces.

A second aspect of the tensor product structure that must be explained is how projectors work. Projectors enable us to express the influence of context, and how transition probabilities and weights are calculated. Suppose we consider a context $e^S \in {\cal M}_S$ of the first concept $S$, represented by a projection operator $P_e^S$ of the Hilbert space ${\cal H}^S$. This context $e^S$ can be considered as a context of the compound $S \otimes T$ of the two concepts $S$ and $T$, and will then be represented by the projection operator $P_e^S \otimes 1^T$, where $1^T$ is the unit operator on ${\cal H}^T$. If we have a context $e^S \in {\cal M}^S$ of the first concept $S$ and a context
$e^T \in {\cal M}^T$ of the second concept $T$, represented respectively by projection operators $P_e^S$ and $P_e^T$, then $P_e^S \otimes P_e^T$ represents the context $e^S
\otimes e^T$ of the compound concept $S \otimes T$.
We have
\begin{equation} \label{eq:projectionstensorproduct}
P_e^S \otimes P_e^T (|x^S\rangle \otimes |x^T\rangle) = P_e^S|x^S\rangle \otimes P_e^T|x^T\rangle
\end{equation}
The transition probabilities and weights are calculated using the following formulas in the tensor product
\begin{equation}
\langle x^S \otimes x^T| y^S \otimes y^T\rangle = \langle x^S|y^S\rangle \langle x^T|y^T\rangle \quad {\rm and} \quad Tr (A^S \otimes A^T) = Tr A^S \cdot Tr A^T
\end{equation}
A third aspect of the tensor product is the reduced states. If the compound quantum entity $S \otimes T$ is
in a nonproduct state $|z\rangle \in {\cal H}^S \otimes {\cal H}^T$ of the tensor product Hilbert space of the two Hilbert spaces ${\cal H}^S$ and
${\cal H}^T$ of the sub-entities, then it is not obvious what states the sub-entities are in, because there are no vectors $|x^S\rangle \in {\cal H}^S$ and $|x^T\rangle \in {\cal H}^T$ such that $|z\rangle = |x^S\rangle \otimes |x^T\rangle$. This means that we can say with certainty that for such a nonproduct state $|z\rangle$, the sub-entities cannot be in pure states. It can be proven in general that the sub-entities are in density states, and these density states are called the reduced states. We do not give the mathematical construction since we only need to calculate the reduced states in specific cases, and refer to \cite{jauch01}, 11-7, for a general definition and derivation of the reduced states.

\subsection{Combining Pet and Fish}
In this section we use the quantum formalism to describe how the concepts `pet' and `fish' combine, and see that the `pet fish problem' \cite{oshersonsmith01,oshersonsmith02,hampton01,fodor01,fodorlepore01} finds a natural solution (see \cite{aertsgabora02} for a presentation of the pet fish problem).

We first have to build the quantum model for the concept `fish', and then combine this, using the tensor product, with the quantum model for `pet'. To provide the necessary data, another experiment was performed, using the same subjects and data acquisition methods as for the experiment in \cite{aertsgabora02}. Subjects were asked to rate the frequency of appearance of different exemplars of `fish' under two contexts: 
\begin{equation}
e_{30}^{fish}, {\rm `The\ fish\ is\ a\ pet'}
\end{equation}
and the unity context $1^{fish}$. We denote the ground state of `fish' by $\hat{p}^{fish}$ and the state under context
$e_{30}^{fish}$ by $p_{30}^{fish}$. The results are presented in Table
\ref{tab:fishratingexemplars}. We note a similar effect than observed previously for the
concept `pet'. For example, the weights of {\it goldfish} and {\it guppy} are greater under context $e_{30}^{fish}$ than for the ground state
under the unity context $1^{fish}$, while the weights of all other exemplars are lower.

\begin{table}[h]
\begin{center}\small
\begin{tabular}{|l||l|l|l|l|} \hline 
\multicolumn{1}{|c||}{exemplar} & \multicolumn{2}{|c|}{$e_{30}^{fish}$} & \multicolumn{2}{|c|}{$1^{fish}$} \\
\hline   
& rate & freq & rate & freq \\ \hline 
{\it trout}  & 0.54 & 0.02 & 4.67 & 0.09 \\ \hline
{\it shark}  & 0.51 & 0.02 & 4.37 & 0.09 \\ \hline
{\it whale}  & 0.15 & 0.01 & 3.36   & 0.07 \\ \hline
{\it dolphin} & 0.91 & 0.04 & 3.72   & 0.07 \\ \hline
{\it pike}  & 0.37 & 0.01 & 2.94   & 0.05 \\ \hline
{\it goldfish}  & 6.73 & 0.40 & 5.19   & 0.10 \\ \hline
{\it ray}  & 0.27 & 0.01 & 3.10    & 0.06 \\ \hline
{\it tuna}  & 0.19 & 0.01 & 4.57  & 0.09 \\ \hline
{\it barracuda}  & 0.40   & 0.01 & 1.53   & 0.03 \\ \hline
{\it mackerel}  & 0.19 & 0.01 & 3.47    & 0.07 \\ \hline
{\it herring}  & 0.22 & 0.01 & 4.46 & 0.09 \\ \hline
{\it guppy}  & 6.60 & 0.39 & 4.10   & 0.08 \\ \hline
{\it plaice}  & 0.22 & 0.01 & 3.56 & 0.07 \\ \hline
{\it carp}  & 1.21 & 0.05 & 3.21   & 0.06 \\ \hline
\end{tabular}
\end{center}
\caption{Frequency ratings of different exemplars of the concept `fish' under two contexts} \label{tab:fishratingexemplars}
\end{table}
Let us call $X^{fish}$ the set of basic contexts and $U^{fish}$ the set of basic states that we consider for the concept `fish'. 
We introduce the states and contexts corresponding to the different exemplars that we have considered in the experiment in Table
\ref{tab:statescontextexemplars02}. So, for example, the context $e_{34}^{fish}$ is the context `The fish is a dolphin' and the state $p_{40}^{fish}$ is
the state of `fish' which is the ground state $\hat{p}^{fish}$ under the context $e_{40}^{fish}$, `The fish is a mackerel'. 
\begin{table}[h] 
\begin{center}
\small\begin{tabular}{|l|l|l|l|l|}
\hline
\multicolumn{1}{|c|}{exemplar} &  \multicolumn{1}{|c|}{context} & \multicolumn{1}{|c|}{state} \\
\hline
{\it trout}  & $e_{31}^{fish}$ & $p_{31}^{fish}$\\
\hline
{\it shark}  & $e_{32}^{fish}$ & $p_{32}^{fish}$ \\
\hline
{\it whale}   & $e_{33}^{fish}$ & $p_{33}^{fish}$ \\ 
\hline
{\it dolphin}  & $e_{34}^{fish}$ & $p_{34}^{fish}$ \\
\hline
{\it pike}  & $e_{35}^{fish}$ & $p_{35}^{fish}$  \\
\hline
{\it goldfish}  & $e_{36}^{fish}$ & $p_{36}^{fish}$ \\
\hline
{\it ray}  & $e_{37}^{fish}$ & $p_{37}^{fish}$ \\
\hline
{\it tuna}  & $e_{38}^{fish}$ & $p_{38}^{fish}$ \\
\hline
{\it barracuda}  & $e_{39}^{fish}$ & $p_{39}^{fish}$ \\
\hline
{\it mackerel}  & $e_{40}^{fish}$ & $p_{40}^{fish}$  \\
\hline
{\it herring}  & $e_{41}^{fish}$ & $p_{41}^{fish}$ \\
\hline
{\it guppy}  & $e_{42}^{fish}$ & $p_{42}^{fish}$  \\
\hline
{\it plaice}  & $e_{43}^{fish}$ &$p_{43}^{fish}$ \\
\hline
{\it carp}  & $e_{44}^{fish}$ & $p_{44}^{fish}$ \\
\hline
\end{tabular}
\end{center}
\caption{The states and contexts connected to the exemplars of the concept `fish' that we considered} \label{tab:statescontextexemplars02}
\end{table}
Further we introduce
\begin{equation}
E_i^{fish} = \{u\ \vert\ u \le e_i^{fish}, u \in X^{fish}\} \quad {\rm and} \quad X_{ij}^{fish} = \{u\ \vert\ u \le e_i^{fish} \wedge e_j^{fish}, u \in X^{fish}\}
\end{equation}
where $E_i^{fish}$ is the set of basic contexts that is stronger or equal to $e_i^{fish}$ and $X_{ij}^{fish}$ the set of basic contexts that is stronger
or equal to
$e_i^{fish}
\wedge e_j^{fish}$. We have $X_{ij}^{fish} = E_i^{fish} \cap E_j^{fish}$. Suppose that $m$ is the total number of basic contexts. Let us denote by
$m(X_{ij}^{fish})$ the number of basic contexts contained in
$X_{ij}^{fish}$ and by
$m(E_i^{fish})$ the number of basic contexts contained in
$E_i^{fish}$.
We choose $m(X_{ij}^{fish})$ and $m(E_i^{fish})$ as in Table \ref{tab:fishbasiccontexts}.
\begin{table}
\begin{center}
\small\begin{tabular}{|l||l|l|} \hline 
\multicolumn{1}{|c||}{exemplar} & \multicolumn{1}{|c|}{$e_{30}^{fish}$} & \multicolumn{1}{|c|}{$1^{fish}$} \\
\hline   
& $m(e_{30}^{fish}) = 100$  & $m = 408$   \\ \hline 
{\it trout}  & $m(X_{31,1}^{fish}) = 2$   & $m(E_{31}^{fish}) = 36$   \\
\hline 
{\it shark}  & $m(X_{32,1}^{fish}) = 2$  & $m(E_{32}^{fish}) = 36$    \\ \hline 
{\it whale}   & $m(X_{33,1}^{fish}) = 1$   & $m(E_{33}^{fish}) = 28$  \\ \hline 
{\it dolphin}  & $m(X_{34,1}^{fish}) = 4$   & $m(E_{34}^{fish}) = 28$   \\ \hline 
{\it pike}  & $m(X_{35,1}^{fish}) = 1$   & $m(E_{35}^{fish}) = 20$   \\ \hline 
{\it goldfish}  & $m(X_{36,1}^{fish}) = 40$    &  $m(E_{36}^{fish}) = 40$    \\
\hline 
{\it ray}  & $m(X_{37,1}^{fish}) = 1$    & $m(E_{37}^{fish}) = 24$   \\ \hline 
{\it tuna}   & $m(X_{38,1}^{fish}) = 1$   & $m(E_{38}^{fish}) = 36$   \\
\hline 
{\it barracuda}    & $m(X_{39,1}^{fish}) = 1$    & $m(E_{39}^{fish}) = 12$    \\
\hline 
{\it mackerel}  & $m(X_{40,1}^{fish}) = 1$    & $m(E_{40}^{fish}) = 28$   \\ \hline 
{\it herring}   & $m(X_{41,1}^{fish}) = 1$   & $m(E_{41}^{fish}) = 36$    \\
\hline 
{\it guppy}  & $m(X_{42,1}^{fish}) = 39$    & $m(E_{42}^{fish}) = 32$   \\ \hline  
{\it plaice}   & $m(X_{43,1}^{fish}) = 1$  & $m(E_{43}^{fish}) = 28$    \\
\hline  {\it carp}  & $m(X_{44,1}^{fish}) = 5$    & $m(E_{44}^{fish}) = 24$ \\ \hline
\end{tabular}
\end{center}
\caption{Choice of the distribution of the different types of basic contexts for the concept `fish'} \label{tab:fishbasiccontexts}
\end{table}
For the quantum model of the concept `fish', we consider a Hilbert space $\compl^m$ of 408 dimensions.

Let us construct the quantum model for the concept `fish'. 
Each basic context $u \in X^{fish}$ is represented by a
projector $|u\rangle \langle u|$, where $|u\rangle \in \compl^m$ is a unit vector, and such that $B^{fish} = \{|u\rangle\ \vert u \in
X^{fish}\}$ is an orthonormal base of the Hilbert space $\compl^m$. The basic state corresponding to the basic context $u$ is represented
by the vector
$|u\rangle$. The ground state
$\hat{p}^{fish}$ of the concept `fish' is represented by the unit vector
$|x_{\hat{p}}^{fish}\rangle$, superposition of the base states $B^{fish} = \{|u\rangle\ \vert\ u \in X^{fish}\}$ using the following expression:
\begin{equation} \label{eq:stateforfish}
|x_{\hat{p}}^{fish}\rangle = \sum_{u \in X^{fish}} {1 \over \sqrt{408}} |u\rangle
\end{equation}
Hence if the concept `fish' is in its ground state $\hat{p}^{fish}$ there is a probability of 1/408 that one of the basic states $u \in U^{fish}$, under contexts $u \in X^{fish}$, is elicited. This means that for `fish' in its ground state, the probability that a basic state gets elicited corresponding to a context contained in $E_i^{fish}$ is given by $m(E_i^{fish})/408$, where $m(E_i^{fish})$ is given in Table \ref{tab:fishbasiccontexts}. A straightforward calculation proves that this gives exactly the weights in Table \ref{tab:fishratingexemplars}. Let us look at some examples. Following
Table \ref{tab:fishbasiccontexts}, in 20 of the 408 basic contexts, the fish is a pike. This means that the weight of pike in the ground state of `fish' is 20/408 = 0.05, which indeed corresponds to what we find in Table
\ref{tab:fishratingexemplars} for pike. In 28 of the 408 basic contexts, the fish is a dolphin. Hence the weight of dolphin in the ground state of `fish' is 28/408 = 0.07, as can be found in Table \ref{tab:fishratingexemplars}. In 32 of the 408 basic contexts, the
fish is a {\it guppy}, thus the weight for {\it guppy} is 32/408 = 0.08, as in Table \ref{tab:fishratingexemplars}.

Now consider the concept `fish' and the context $e_{30}^{fish}$,
`The fish is a pet'. The context
$e_{30}^{fish}$ is represented by the projection operator $P_{e_{30}}^{fish}$ given by
\begin{equation}
P_{e_{30}}^{fish} = \sum_{u \in E_{30}^{fish}} |u\rangle \langle u|
\end{equation}
where $E_{30}^{fish}$ is the set of basic contexts of `fish' that is stronger than or equal to $e_{30}^{fish}$, hence $E_{30}^{fish} = \{u\ \vert\ u \le e_{30}^{fish}, u \in X^{fish}\}$. Let us calculate the new state $|x_{p_{30}}^{fish}\rangle$ that $|x_{\hat{p}}^{fish}\rangle$ changes to under the influence of $e_{30}^{fish}$.
Following (\ref{eq:contextchangestate}) we have
\begin{equation}
|x_{p_{30}}^{fish}\rangle = {P_{e_{30}}^{fish} |x_{\hat{p}}^{fish}\rangle \over \sqrt{\langle x_{\hat{p}}^{fish}| P_{e_{30}^{fish}}
|x_{\hat{p}}^{fish}\rangle}}
\end{equation}
We have
\begin{equation}
P_{e_{30}}^{fish} |x_{\hat{p}}^{fish}\rangle = \sum_{u \in E_{30}^{fish}} |u\rangle \langle u|x_{\hat{p}}^{fish}\rangle = \sum_{u \in E_{30}^{fish}} {1
\over
\sqrt{408}}|u\rangle
\end{equation}
and
\begin{equation}
\langle x_{\hat{p}}^{fish}| P_{e_{30}}^{fish} |x_{\hat{p}}^{fish}\rangle = \sum_{u \in E_{30}^{fish}} \langle x_{\hat{p}}^{fish}|u\rangle \langle u|
x_{\hat{p}}^{fish}\rangle =
\sum_{u
\in E_{30}^{fish}} |\langle x_{\hat{p}}^{fish}|u\rangle|^2 = \sum_{u \in E_{30}^{fish}}{1 \over 408} = {100 \over 408}
\end{equation}
This gives
\begin{equation} \label{eq:statefishispet}
|x_{p_{30}}^{fish}\rangle = \sum_{u \in E_{30}^{fish}} {1 \over \sqrt{100}} |u\rangle
\end{equation}

\subsection{The Compound Pet $\otimes$ Fish} \label{sec:compound}
The compound of the concepts `pet' and `fish', denoted `pet $\otimes$ fish', is described in the space $\compl^n \otimes \compl^m$. A specific combination does not correspond to the totality of the new concept `pet $\otimes$ fish', but rather to a subset of it. For example, the combination `a pet and a fish' is one subset of states of `pet $\otimes$ fish', and the combination `pet fish' is another. As we will see, `a pet and a fish' corresponds to a subset containing only product states of `pet $\otimes$ fish', while `pet fish' corresponds to a subset containing entangled states of `pet $\otimes$ fish'. Let us analyze what is meant by different possible states of the compound `pet $\otimes$ fish' of the concepts `pet' and `fish', hence vectors or density operators of the tensor product Hilbert space $\compl^n \otimes \compl^m$. 

The first state we consider is $\hat{p}^{pet} \otimes \hat{p}^{fish}$, the tensor product of the ground state $\hat{p}^{pet}$ of `pet' and the ground state $\hat{p}^{fish}$ of `fish' , which is represented in $\compl^n \otimes \compl^m$ by the vector $|x_{\hat{p}}^{pet}\rangle \otimes |x_{\hat{p}}^{fish}\rangle$. This state is a good representation of the conceptual combination `pet and fish', because for `pet and fish', contexts can act on `pet', or on `fish', or both, and they act independently. More concretely, consider the context $e_1^{pet}$, `The pet is chewing a bone' acting on the concept `pet'. This context, then written like $e_1^{pet} \otimes 1^{fish}$, can also act on the `pet' sub-concept of `pet $\otimes$ fish'. Then this will just change the ground state $\hat{p}^{pet}$ of `pet' to state $p_1^{pet}$, and the ground state $\hat{p}^{fish}$ of the `fish' sub-concept of `pet $\otimes$ fish' will not be influenced. This is exactly the kind of change that the state represented by $|x_{\hat{p}}^{pet}\rangle \otimes |x_{\hat{p}}^{fish}\rangle$ entails.

Hence
\begin{eqnarray} \label{eq:producttransforms01}
\hat{p}^{pet} \buildrel e_1^{pet} \over \longmapsto p_1^{pet} \quad &\Rightarrow& \quad \hat{p}^{pet} \otimes \hat{p}^{fish} \buildrel e_1^{pet} \otimes
1^{fish}
\over
\longmapsto p_1^{pet}
\otimes
\hat{p}^{fish}
\\ &\Rightarrow& \quad \hat{p}^{pet} \otimes p_{30}^{fish} \buildrel e_1^{pet} \otimes 1^{fish} \over \longmapsto p_1^{pet} \otimes p_{30}^{fish}
\end{eqnarray}
Similarly, a context that only works on the concept `fish', can work on the `fish' sub-concept of `pet $\otimes$ fish', and in this case will not
influence the state of `pet'. Hence
\begin{eqnarray} \label{eq:producttransforms02}
\hat{p}^{fish} \buildrel e_{30}^{fish} \over \longmapsto p_{30}^{fish} \quad &\Rightarrow& \quad \hat{p}^{pet} \otimes \hat{p}^{fish} \buildrel 1^{pet}
\otimes e_{30}^{fish}
\over \longmapsto \hat{p}^{pet} \otimes p_{30}^{fish} \\ 
&\Rightarrow& \quad p_i^{pet} \otimes \hat{p}^{fish} \buildrel 1^{pet} \otimes e_{30}^{fish} \over \longmapsto p_i^{pet} \otimes p_{30}^{fish}
\end{eqnarray}
Another state to consider is $p_6^{pet} \otimes p_{30}^{fish}$, represented by the vector $|x_{p_6}^{pet}\rangle \otimes
|x_{p_{30}}^{fish}\rangle$. This is a state where the `pet' is a `fish' and the `fish' is a `pet'; hence perhaps this state faithfully represents `pet fish'. How can we check this? We begin by verifying different frequencies of exemplars and weights of properties in this state, and seeing whether the guppy effect, described in Section 2.1 of \cite{aertsgabora02}, is predicted by the model. (\ref{eq:weightgoldfishpetfish}) gives the calculation for the weight of the exemplar {\it goldfish} for the concept `pet' in state $p_6^{pet}$. Now we calculate the weight for the exemplar {\it goldfish} for the compound concept `pet $\otimes$ fish' in state $p_6^{pet} \otimes p_{30}^{fish}$. Following the quantum mechanical rules outlined in
(\ref{eq:projectionstensorproduct}) we need to apply the projector $P_{e_{18}}^{pet} \otimes 1^{fish}$ on the vector $|x_{p_6}^{pet}\rangle
\otimes |y_{p_{30}}^{fish}\rangle$, and use it in the quantum formula (\ref{eq:contextchangestateprobability}). This gives:
\begin{eqnarray} \label{eq:weightgoldfishpetfishtensor}
\mu(p_{18}^{pet} \otimes p_{30}^{fish}, e_{18}^{pet} \otimes 1^{fish}, p_6^{pet} \otimes p_{30}^{fish}) &=& (\langle x_{p_6}^{pet}| \otimes \langle
x_{p_{30}}^{fish}|) (P_{e_{18}}^{pet}
\otimes 1^{fish}) (|x_{p_6}^{pet}\rangle
\otimes |x_{p_{30}}^{fish}\rangle)
\\ &=& \langle x_{p_6}^{pet}|  P_{e_{18}}^{pet} |x_{p_6}^{pet}\rangle   \langle x_{p_{30}}^{fish}| x_{p_{30}}^{fish}\rangle = \langle x_{p_6}^{pet}|  P_{e_{18}}^{pet} |x_{p_6}^{pet}\rangle \\
&=& {48 \over 100} = 0.48 \label{eq:outcomeguppyeffect01}
\end{eqnarray}
This means that the weight of the exemplar {\it goldfish} of the sub-concept `pet' of the compound `pet $\otimes$ fish' in the product state $p_6^{pet}
\otimes p_{30}^{fish}$ (the state that represents a `pet $\otimes$ fish' that is a pet and a fish), is equal to the weight of the exemplar {\it goldfish} of the concept `pet' in state $p_6^{pet}$ (the state that represents a pet that is a fish). This is not surprising; it simply means that the tensor product in its simplest type of state, the product state, takes over the weights that were there already for the separate sub-concepts. The guppy effect, identified previously in states $p_6^{pet}$ of the concept `pet' and $p_{30}^{fish}$ of the concept `fish', remains there in this combination of pet and fish described by this product state $p_6^{pet} \otimes p_{30}^{fish}$. Indeed, we can repeat the calculation of (\ref{eq:weightgoldfishpetfishtensor}) on the product state of the ground states---hence the state $\hat{p}^{pet} \otimes \hat{p}^{fish}$---and find
\begin{eqnarray} \label{eq:weightgoldfishpetfishtensor02}
\mu(p_{18}^{pet} \otimes \hat{p}^{fish}, e_{18}^{pet} \otimes 1^{fish}, \hat{p}^{pet} \otimes \hat{p}^{fish}) &=& (\langle x_{\hat{p}}^{pet}| \otimes
\langle x_{\hat{p}}^{fish}|) (P_{e_{18}}^{pet}
\otimes 1^{fish}) (|x_{\hat{p}}^{pet}\rangle
\otimes |x_{\hat{p}}^{fish}\rangle)
\\ &=& \langle x_{\hat{p}}^{pet}|  P_{e_{18}}^{pet} |x_{\hat{p}}^{pet}\rangle   \langle x_{\hat{p}}^{fish}| x_{\hat{p}}^{fish}\rangle = \langle x_{\hat{p}}^{pet}|  P_{e_{18}}^{pet} |x_{\hat{p}}^{pet}\rangle \\
&=& {140 \over 1400} = 0.10 \label{eq:outcomeguppyeffect02}
\end{eqnarray}
We see that the weight of {\it goldfish} for the sub-concept `pet' of the compound `pet $\otimes$ fish' equals the weight of {\it goldfish} for the concept `pet' in the ground state $\hat{p}^{pet}$. The difference between (\ref{eq:outcomeguppyeffect01}) and (\ref{eq:outcomeguppyeffect02}) is the guppy effect in our theory of the compound `pet $\otimes$ fish'. It should be stated in the following way. The weight of {\it goldfish} of the concept `pet' equals 0.10 if `pet' is in its ground state, and equals 0.48 if `pet' is in a state under the context `The pet is a fish'. This is the pre-guppy effect identified by introducing contexts for the description of one concept, namely `pet'. When `pet' combines with `fish' we get the concept `pet $\otimes$ fish'. Now the {\it guppy} effect manifests in the following way. The weight of
{\it goldfish} for `pet' as a sub-concept of `pet $\otimes$ fish' equals 0.10 if the state of `pet $\otimes$ fish' is such that we have `a pet and a fish' in
the state `a pet \ldots and \ldots a fish' (without necessarily the pet being a fish and the fish being a pet, this is the product state of the two
ground states, hence
$\hat{p}^{pet}
\otimes
\hat{p}^{fish}$). The weight of {\it goldfish} for `pet' as a sub-concept of `pet' $\otimes$ fish' equals 0.48 if the state of `pet $\otimes$ fish' is such that we have `a pet and a fish' in a state where the pet is a fish and the fish is a pet (this is the product state $p_6^{pet} \otimes p_{30}^{fish}$). So we get the guppy effect in the combination of the concepts `pet' and `fish'. But does this mean that the state $p_6^{pet} \otimes p_{30}^{fish}$ describes a `pet fish'? The weights of exemplars seem to indicate this, but there is still something fundamentally wrong. Look at formula (\ref{eq:weightgoldfishpetfishtensor}). It reads $\mu(p_{18}^{pet} \otimes p_{30}^{fish}, e_{18}^{pet} \otimes 1^{fish}, p_6^{pet} \otimes p_{30}^{fish})$. This means that under the influence of context $e_{18}^{pet} \otimes 1^{fish}$ state $p_6^{pet} \otimes p_{30}^{fish}$ changes to state $p_{18}^{pet} \otimes p_{30}^{fish}$. The state
$p_6^{pet} \otimes p_{30}^{fish}$ is a product state of the compound `pet
$\otimes$ fish' where the pet is a fish and the fish is a pet. But if `pet' as sub-concept of the compound collapses to {\it goldfish} (this is the state transformation $p_6^{pet} \longmapsto p_{18}^{pet}$), we see that $p_{30}^{fish}$ remains unchanged in the collapse translated to the compound (we have there $p_6^{pet} \otimes p_{30}^{fish} \longmapsto p_{18}^{pet} \otimes p_{30}^{fish}$). This means that the context `The pet is a goldfish' causes `pet' as a sub-concept to collapse to {\it goldfish}, but leaves `fish' as a sub-concept unchanged. The end state after the collapse is $p_{18}^{pet} \otimes p_{30}^{fish}$, which means `a goldfish and a fish' (pet has become goldfish, but fish has remained fish). We could have expected this, because the rules of the tensor product tell us exactly that product states behave this way. Their rules are given in symbolic form in (\ref{eq:producttransforms01}) and
(\ref{eq:producttransforms02}). Product states describe combined concepts that remain independent, {\it i.e.} the concepts are combined in such a way that the influence of a context on one of the sub-concepts does not influence the other sub-concept. That is why, as mentioned previously, the product states describe the combination with the `and' between the concepts; hence `pet and fish'. Then what does the product state $p_6^{pet} \otimes p_{30}^{fish}$ describe? It describes the situation where the pet is a fish, and the fish is a pet: hence two `pet fish' and not one! And indeed, the mathematics shows us this subtlety. If for two `pet fish', one collapses to {\it goldfish}, there is no reason at all that the other also collapses to {\it goldfish}. It might for example be {\it goldfish} and {\it guppy}. So to clarify what we are saying here, a possible instance of state $p_6^{pet} \otimes p_{30}^{fish}$ of the compound `pet $\otimes$ fish' is `a goldfish and a guppy'. Now we can see why this state $p_6^{pet} \otimes p_{30}^{fish}$ gives numerical indication of a guppy effect. But we did not really find the guppy effect, for the simple reason that we did not yet identify the state that describes `pet fish' (one unique living being that is a `pet' and a `fish'). It is here that one of the strangest and most sophisticated of all quantum effects comes in, namely entanglement.

\subsection{The `Pet Fish' as a Quantum Entangled State} \label{sec:petfishentangled}
Consider the context
\begin{equation}
e_{45}, {\rm `The\ pet\ swims\ around\ the\ little\ pool\ where\ the\ fish\ is\ being\ fed\ by\ the\ girl'}
\end{equation}
This is a context of `pet' as well as of `fish'. It is possible to consider a big reservoir of contexts that have not yet been
classified as a context of a specific concept. We denote this reservoir ${\cal M}$. This means concretely that ${\cal M}^{pet} \subset {\cal M}$ and
${\cal M}^{fish} \subset {\cal M}$. Let is denote ${\cal M}^{pet,fish}$ the set of contexts that are contexts of `pet' and also contexts of `fish'.
Amongst the concrete contexts that were considered in this paper, there are seven that are elements of ${\cal M}^{pet,fish}$, namely
\begin{equation}
e_6, e_{18}, e_{21}, e_{30}, e_{36}, e_{42}, e_{45} \in {\cal M}^{pet,fish}
\end{equation}
We denote $X^{pet,fish}$ the set of basic contexts that are contexts of `pet' as well as contexts of `fish'. We have
\begin{equation}
E_6^{pet} \subset X^{pet,fish} \quad {\rm and} \quad E_{30}^{fish} \subset X^{pet,fish}
\end{equation}
and to model the concept `pet fish' we make the hypothesis that $E_6^{pet} = E_{30}^{fish} = E^{pet,fish}$, namely that the basic contexts of `pet' where
the pet is a fish are the same as the basic contexts of `fish' where the fish is a pet. It is not strictly necessary to hypothesize that these two
sets are equal. It is sufficient to make the hypothesis that there is a subset of both that contains the basic contexts of `pet' as well as of `fish'
that are also basic context of a pet that is a fish.

We have now everything that is necessary to put
forth the entangled state that describes `pet fish'. It is the following state
\begin{equation} \label{eq:entangledstate}
|s\rangle = \sum_{u \in E^{pet,fish}} {1 \over
\sqrt{100}} |u\rangle
\otimes |u\rangle
\end{equation}
We claim that this vector represents the state of `pet $\otimes$ fish' that corresponds to the conceptual combination `pet fish'. Let us
denote it with the symbol $s$.

Now we have to verify what the states of the sub-concepts `pet' and `fish' are if the compound concept `pet $\otimes$ fish' is in
the state $s$ represented by $|s\rangle$. Hence let us calculate the reduced states for both `pet' and `fish' of the state
$|s\rangle$. As explained in Section \ref{sec:tensorproduct}, for a non-product vector, the reduced states are density operators, not vectors. We
first calculate the density operator corresponding to $|s\rangle \in \compl^n \otimes \compl^m$. This is given by
\begin{equation}
|s\rangle \langle s| = (\sum_{u \in E^{pet,fish}} {1 \over \sqrt{100}} |u\rangle \otimes |u\rangle)
(\sum_{v
\in E^{pet,fish}} {1 \over \sqrt{100}} \langle v| \otimes \langle v|) 
= \sum_{u, v \in E^{pet,fish}} {1 \over 100} |u\rangle \langle v| \otimes |u\rangle \langle
v|
\end{equation}
We find the two reduced density operators by exchanging one of the two products $|u\rangle \langle v|$ by
the inproduct $\langle u| v\rangle$. Taking into account
that $\langle u| v\rangle = \delta(u, v)$, we have
\begin{equation}
|s\rangle \langle s|^{pet} = \sum_{u \in E^{pet,fish}} {1 \over 100} |u\rangle \langle u| \quad {\rm and} \quad |s\rangle \langle s|^{fish} = \sum_{u \in E^{pet,fish}} {1 \over 100} |u\rangle \langle u|
\end{equation}
as reduced states for `pet' and `fish' respectively. It can be proven that these reduced states behave exactly like the states $p_6^{pet}$ and
$p_{30}^{fish}$ respectively. This means that for influences of contexts and weights of properties limited to one of the two sub-concepts `pet'
or `fish', the state $|s\rangle$ behaves exactly as would the product state $|x_{p_6}^{pet}\rangle \otimes |x_{p_{30}}^{fish}\rangle$. This means
that as far as the weights of exemplars and properties are concerned, we find the values that have been calculated for the state $|x_{p_6}^{pet}\rangle
\otimes |x_{p_{30}}^{fish}\rangle$ in the previous section when the compound concept `pet
$\otimes$ fish' is in the entangled state $|s\rangle$.

Let us now see how the state $|s\rangle$ changes under the influence of the context $e_{18}^{pet} \otimes 1^{fish}$, `The pet is a goldfish'
of the concept `pet'. We have 
\begin{equation}
P_{e_{18}}^{pet} \otimes 1^{fish}= \sum_{u \in E_{18}^{pet}} |u\rangle \langle u| \otimes 1
\end{equation}
where $E_{18}^{pet} = \{u\ \vert\ u \le e_{18}^{pet}, u \in X\}$. Hence the changed state of $s$ under the influence of context $e_{18}^{pet} \otimes 1^{fish}$---let us denote it $s'$---is given by
\begin{eqnarray}
|s'\rangle &=& (P_{e_{18}}^{pet} \otimes 1^{fish}) |s\rangle = \sum_{u \in E_{18}^{pet}} \sum_{v \in E_6^{pet}} |u\rangle \langle u| \otimes 1 {1 \over
\sqrt{100}} |v\rangle
\otimes |v\rangle \\ 
&=& \sum_{u \in E_{18}^{pet}} \sum_{v \in E_6^{pet}} {1 \over \sqrt{100}} \langle u|v\rangle |u\rangle \otimes |v\rangle = \sum_{u \in E_{18}^{pet}}
\sum_{v \in E_6^{pet}} {1
\over \sqrt{100}} \delta(u, v) |u\rangle \otimes |v\rangle \\ 
&=& \sum_{u \in E_{18}^{pet} \cap E_6^{pet}} {1 \over \sqrt{100}} |u\rangle \otimes |u\rangle
\end{eqnarray}
Calculating the reduced density states gives
\begin{equation}
|s'\rangle \langle s'|^{pet} = \sum_{u \in E_{18}^{pet} \cap E_6^{pet}} {1 \over 100} |u\rangle \langle u| \quad {\rm and} \quad |s'\rangle \langle s'|^{fish} = \sum_{u \in E_{18}^{pet} \cap E_6^{pet}} {1 \over 100} |u\rangle \langle u|
\end{equation}
The reduced state $|s'\rangle \langle s'|^{pet}$ with respect to the concept `pet' is
the state of `pet' under the context $e_6^{pet}$, `The pet is a fish', and the context
$e_{18}^{pet}$, `The pet is a goldfish'. This is what we would have expected in any case, because indeed the context $e_{18}^{pet}$, 
influences `pet' alone and not `fish'. However, the reduced state $|s'\rangle \langle s'|^{fish}$ with respect to the concept `fish' after the change
provoked by the context
$e_{18}^{pet}$, `is a goldfish', that only influences the concept `pet' directly, is also a state of `fish' under the context `is a pet' and
under the context `is a goldfish'. This means that if for `pet fish' the pet becomes a {\it goldfish}, then also for `fish' the fish becomes a {\it goldfish}.
This is exactly what is described by the entangled state $|s\rangle$ of the tensor product space given in (\ref{eq:entangledstate}).

\subsection{Combining Concepts in Sentences} \label{sec:sentences}
In this section we apply our formalism to model more than two combinations of concepts. Consider a simple archetypical sentence containing a subject, and object and a predicate connecting both: `The cat eats the food'. Three concepts `cat', `eat' and `food' are involved: two nouns and one verb. We want to show that it is possible to represent this sentence as an entangled state of the
compound concept `cat $\otimes$ eat $\otimes$ food'.

We introduce the SCOPs of `cat', `eat' and `food', $(\Sigma^{cat}, {\cal M}^{cat}, {\cal L}^{cat}, \mu^{cat}, \nu^{cat})$,
$(\Sigma^{eat}, {\cal M}^{eat},$ ${\cal L}^{eat}, \mu^{eat}, \nu^{eat})$ and $(\Sigma^{food}, {\cal M}^{food}, {\cal L}^{food}, \mu^{food},
\nu^{food})$. ${\cal M}$ is the reservoir of contexts that have not been decided to be relevant for a specific concept, hence ${\cal M}^{cat}
\subset {\cal M}, {\cal M}^{eat} \subset {\cal M}$ and ${\cal M}^{food} \subset {\cal M}$. We choose Hilbert spaces ${\cal H}^{cat}$, ${\cal H}^{eat}$ and ${\cal H}^{food}$ to represent respectively the concepts `cat', `eat' and
`food'. Then we construct the tensor product Hilbert space ${\cal H}^{cat} \otimes {\cal H}^{eat} \otimes {\cal H}^{food}$ to represent the
compound concept `cat
$\otimes$ eat $\otimes$ food'. Consider the three ground states $|x_{\hat{p}}^{cat}\rangle \in {\cal H}^{cat}$, $|x_{\hat{p}}^{eat}\rangle \in
{\cal H}^{eat}$ and
$|x_{\hat{p}}^{food}\rangle
\in
{\cal H}^{food}$ of respectively `cat', `eat' and `food'. The product state $|x_{\hat{p}}^{cat}\rangle \otimes |x_{\hat{p}}^{eat}\rangle \otimes
|x_{\hat{p}}^{food}\rangle \in {\cal H}^{cat} \otimes {\cal H}^{eat} \otimes {\cal H}^{food}$ represents the conceptual combination `cat and eat and
food'. Although it is
technically the simplest combination, the one described by the product state of the three ground states of each concept apart, it
is rare in everyday life. Indeed, upon exposure to the three concepts `cat' `eat' `food' in a row, the mind seems to be caught in a
spontaneous act of entanglement that generates the sentence `the cat eats the food'. We will come back to this later, because the same phenomenon exists with quantum entities. For the moment, however, let us consider the three concepts `cat', `eats' and `food' connected by the word `and' in a independent, hence non-entangled way; {\it i.e.} 
`cat and eat and food' described by the product state
$|x_{\hat{p}}^{cat}\rangle \otimes |x_{\hat{p}}^{eat}\rangle \otimes |x_{\hat{p}}^{food}\rangle$. Concretely this means that if a specific context
influences the concept `cat', then the concepts `eat' and `food' are not influenced. For example, suppose that the ground state
$|x_{\hat{p}}^{cat}\rangle$ of the concept `cat' is changed by the context
\begin{equation}
e_{46}^{cat}, {\rm `The\ cat\ is\ Felix'}
\end{equation}
into the state
$p_{46}^{cat}$,`The cat is Felix'. If this context $e_{46}^{cat}$ is applied to the compound concept `cat $\otimes$ eat
$\otimes$ food' in the product state
$|x_{\hat{p}}^{cat}\rangle \otimes |x_{\hat{p}}^{eat}\rangle \otimes |x_{\hat{p}}^{food}\rangle$, then the compound concept changes state to
$|x_{p_{46}}^{cat}\rangle \otimes |x_{\hat{p}}^{eat}\rangle \otimes |x_{\hat{p}}^{food}\rangle$
\begin{equation}
|x_{\hat{p}}^{cat}\rangle \otimes |x_{\hat{p}}^{eat}\rangle \otimes |x_{\hat{p}}^{food}\rangle \buildrel e_{46}^{cat} \otimes 1^{eat} \otimes 1^{food}
\over
\longmapsto |x_{p_{46}}^{cat}\rangle \otimes |x_{\hat{p}}^{eat}\rangle \otimes |x_{\hat{p}}^{food}\rangle
\end{equation}
This state expresses `Felix and eat and food' as a state of the compound concept `cat $\otimes$ eat $\otimes$ food'. Can we determine the state of the
compound concept `cat $\otimes$ eat $\otimes$ food' that describes the sentence `The cat eats the food'? Again, as in the case of `pet fish' this will
be an entangled state of the tensor product Hilbert space. Indeed, for the sentence `The cat eats the food', we require that if, for example,
`cat' collapses to `Felix', then also `eat' must collapse to `Felix who eats', and `food' must collapse to `Felix and the food she eats'. This means that the sentence `The cat eats the food' is certainly not described by a products state of the tensor product Hilbert space.
How do we build the correct entangled state? Let us explain this step by step so that we can see how this could work for any arbitrary sentence.

First, we observe that the sentence itself is a context for `cat', `eat' and `food'. Let us call it $e_{47}$, hence
\begin{equation}
e_{47}, {\rm `The\ cat\ eats\ the\ food'}
\end{equation}
We have $e_{47} \in {\cal M}$, but also $e_{47}^{cat} \in {\cal M}^{cat}$, $e_{47}^{eat} \in {\cal M}^{eat}$ and $e_{47}^{food} \in {\cal M}^{food}$.
Now we introduce $E_{47} = \{u\ \vert u \le e_{47}, u \in X\}$
is the set of basic contexts that are stronger than or equal to $e_{47}$.
The entangled state, element of the tensor product Hilbert space ${\cal H}^{cat} \otimes {\cal H}^{eat} \otimes {\cal H}^{food}$, that describes the
sentence `The cat eats the food' is given by
\begin{equation}
|s\rangle = \sum_{u \in E_{47}} {1 \over \sqrt{n(E_{47})}}|u\rangle \otimes |u\rangle \otimes |u\rangle
\end{equation}
where $n(E_{47})$ is the number of basic contexts contained in $E_{47}$.

Let us show that this state describes exactly the entanglement of the sentence `The cat eats the food'. 
We calculate the reduced states of `cat', `eat' and `food' when the compound `cat $\otimes$ eat $\otimes$ food' is in the state $s$ represented by
$|s\rangle$. We
first calculate the density operator corresponding to $|s\rangle$. This is given by
\begin{eqnarray}
|s\rangle \langle s| &=& (\sum_{u \in E_{47}} {1 \over \sqrt{n(E_{47})}} |u\rangle \otimes |u\rangle \otimes |u\rangle)
(\sum_{v
\in E_{47}} {1 \over \sqrt{n(E_{47})}} \langle v| \otimes \langle v| \otimes \langle v|) \\
&=& \sum_{u, v \in E_{47}} {1 \over n(E_{47})} |u\rangle \langle v| \otimes |u\rangle \langle v| \otimes |u\rangle \langle v|
\end{eqnarray}
This gives us
\begin{eqnarray}
&&|s\rangle \langle s|^{cat} = \sum_{u \in E_{47}^{cat}} {1 \over n(E_{47}^{cat})} |u\rangle \langle u| \\
&&|s\rangle \langle s|^{eat} = \sum_{u \in E_{47}^{eat}} {1 \over n(E_{47}^{eat})} |u\rangle \langle u| \\
&&|s\rangle \langle s|^{food} = \sum_{u \in E_{47}^{food}} {1 \over n(E_{47}^{food})} |u\rangle \langle u|
\end{eqnarray}
as reduced states for `cat', `eat' and `food' respectively. These reduced states behave exactly like the states $p_{47}^{cat}$, $p_{47}^{eat}$ and
$p_{47}^{food}$ of respectively `cat', `eat' and `food', when it comes to calculating frequency values of exemplars and applicability values of properties.

Let us now see how the state $|s\rangle$ changes under the influence of the context $e_{46}^{cat} \otimes 1^{eat} \otimes 1^{food}$, `The cat is Felix' of
the concept `cat' as a sub-concept of the compound concept `cat $\otimes$ eat $\otimes$ food'. We have 
\begin{equation}
P_{e_{46}}^{cat} \otimes 1^{eat} \otimes 1^{food} = \sum_{u \in E_{46}^{cat}} |u\rangle \langle u| \otimes 1 \otimes 1
\end{equation}
where $E_{46}^{cat} = \{u\ \vert\ u \le e_{46}^{cat}, u \in X^{cat}\}$. Hence the changed state of $s$ under the influence of context $e_{46}^{cat} \otimes 1^{eat} \otimes 1^{food}$---let us denote it $s'$---is given by
\begin{eqnarray}
|s'\rangle &=& (P_{e_{46}}^{cat} \otimes 1^{eat} \otimes 1^{food}) |s\rangle = \sum_{u \in E_{46}^{cat}} \sum_{v \in E_{47}} |u\rangle \langle u|
\otimes 1 \otimes 1 {1
\over
\sqrt{n(E_{47})}} |v\rangle
\otimes |v\rangle \otimes |v\rangle  \\ 
&=& \sum_{u \in E_{46}^{cat}} \sum_{v \in E_{47}} {1 \over \sqrt{n(E_{47})}} \langle u|v\rangle |u\rangle \otimes |v\rangle \otimes |v\rangle = \sum_{u \in
E_{46}^{cat}}
\sum_{v
\in E_{47}} {1
\over \sqrt{n(E_{47})}} \delta(u, v) |u\rangle \otimes |v\rangle \otimes |v\rangle \\ 
&=& \sum_{u \in E_{46}^{cat} \cap E_{47}} {1 \over \sqrt{n(E_{47})}} |u\rangle \otimes |u\rangle \otimes |u\rangle
\end{eqnarray}
Calculating the reduced density states gives
\begin{eqnarray}
&&|s'\rangle \langle s'|^{cat} = \sum_{u \in E_{46}^{cat} \cap E_{47}^{cat}} {1 \over n(E_{47})} |u\rangle \langle u| \\
&&|s'\rangle \langle s'|^{eat} = \sum_{u \in E_{46}^{cat} \cap E_{47}^{eat}} {1 \over n(E_{47})} |u\rangle \langle u| \\
&&|s'\rangle \langle s'|^{food} = \sum_{u \in E_{46}^{cat} \cap E_{47}^{food}} {1 \over n(E_{47})} |u\rangle \langle u|
\end{eqnarray}
The reduced state $|s'\rangle \langle s'|^{cat}$ with respect to the concept `cat' is
the state of `cat' under the context $e_{46}^{cat} \wedge e_{47}$, `The cat is Felix and the cat eats the food'. This is what we would have expected in
any case, because indeed the context $e_{46}^{cat} \otimes 1^{eat} \otimes 1^{food}$ influences `cat' alone and not `eat' and `food'.
However, the reduced state
$|s'\rangle
\langle s'|^{eat}$ with respect to the concept `eat' after the change provoked by the context
$e_{46}^{cat} \otimes 1^{eat} \otimes 1^{food}$, `The cat is Felix', that only influences `cat' directly, is also a state of `eat' under the
context $e_{46}^{cat} \wedge e_{47}$, `The cat is Felix and the cat eats the food', hence `Felix eats the food'. This means that if for `The cat eats
the food' the `cat` becomes `Felix', then also `eat' becomes `Felix who eats'. A similar phenomenon happens for the concept `food'. The reduced
state $|s'\rangle \langle s'|^{food}$ after the change provoked by the context
$e_{46}^{cat} \otimes 1^{eat} \otimes 1^{food}$, `The cat is Felix', that only influences `cat' directly, is also a state of `food' under the
context $e_{46}^{cat} \wedge e_{47}$, `The cat is Felix and the cat eats the food', hence `Felix eats the food'. This means that if for
`The cat eats the food' the `cat` becomes `Felix', then also `food' becomes `Felix who eats the food'.
The approach that we have put forward in this article can be used to elaborate the vector space models for
representing words that are used in semantic analysis. The tensor product,
and the way that we introduced entangled states to represent sentences, can be used to `solve' the well known `bag of word' problem  (texts are treated as
`bag of words', hence order and syntax cannot be taken into account) as formulated in semantic analysis \cite{aertsczachor01}. In a forthcoming paper we investigate more directly how the quantum structures introduced in \cite{aertsgabora02}, {\it i.e.} the complete orthocomplemented lattice
structure, can be employed in semantic analysis models, and also the relation of our approach with ideas formulated in
\cite{widdow01,widdowspeters01} about quantum logic and semantic analysis.

\subsection{A Quantum Theory of Memory}
In \cite{vonfoerster01} von Foerster develops a theory of memory and hints how a quantum mechanical formalism could be used to formalize the theory. Von Foerster was inspired by how quantum mechanics was introduced in biology. Genes, the carriers of heredity, are described as quantized states of complex molecules. Von Foerster introduces what he calls {\it carriers of elementary impressions}, which he calls {\it mems}, to stress the analogy with {\it genes}, and introduces the notion of {\it impregnation} as an archetypical activation of a carrier by an impression. Such an {\it impregnation of a mem} is formalized as a {\it quantum mechanical excitation} of one energy level of the mem to another energy level of this same mem, in analogy how this happens with a molecule. A molecule in an excited state spontaneously falls back to a lower energy state, and this process is called decay. The decay process of a mem in a high level energy state to a lower level energy state describes the phenomenon of {\it forgetting}. The introduction of the quantum mechanical mechanism of excitation and decay between different energy levels of a mem as the fundamental process of memory, respectively accounting for the learning and the forgetting process, is not developed further in von Foerster's publication. Von Foerster's conviction about the relevance of quantum mechanics to memory comes from his phenomenological study of the dynamics of the forgetting process. Although is not very explicit about this aspect, it can be inferred from his article that in his opinion the physical carrier of the mem is a molecule in the brain, such as a large protein, and that memory is hence stored within a micro-physical entity, entailing quantum structure because of its micro-physical nature. 

The theory of concepts that we have elaborated is in some respects quite different from von Foerster's approach, but in other respects can deliver a possible theoretical background for this approach. Different in the sense that we do not believe that necessary there need to be a micro-physical carrier for the quantum structure identified in SCOP. It is not excluded that the quantum structure is encrypted in a quite unique way in the brain, making use of the possibility to realize quantum structure in the macro-world, without the need of micro-physical entities \cite{aerts08,aerts09,aertsetal14,aertsetal15,aertsetal01,aertsetal13}. On the other hand, if micro-physical entities in the brain serve as carriers of quantum mechanical structure, our SCOP theory could provide specific information about this structure. We can also now clarify the notion of ground state. If a concept is not evoked in any specific kind of way, which is equivalent to it being under the influence of the bath of all type of contexts that can evoke it, we consider it to be in its ground state. Here we align our theory with von Foerster's idea and use the quantum mechanical processes of excitation and decay to point out specific influences of contexts on the state of a concept. If the concept `pet', changes to the state $p_1$ under the influence of context $e_1$, `The pet is chewing a bone', then $p_1$ is an excited state with respect to the ground state $\hat{p}$ of `pet'. The state $p_1$ will spontaneously decay to the ground state $\hat{p}$. We `forget' after a little while the influence of context $e_1$, `The pet is chewing a bone' on the concept `pet' and consider `pet' again in its ground state when a new context arrives that excites it again to another state. The process of excitation and de-excitation or decay, goes on in this way, and constitutes the basic dynamics of a concept in interaction with contexts. This is very much aligned with what von Foerster intuitively had in mind in \cite{vonfoerster01}, and fits completely with a further quantum mechanical elaboration of our SCOP theory of concepts. We can go some steps further in this direction, which is certainly still speculative, but worth mentioning since it shows some of the possible perspectives that can be investigated in future research. If a molecule de-excites (or decays) and collapses to its ground state (or to a lower energy state) it sends out a photon exactly of the amount of energy that equals the difference between the energy of the ground state (the lower energy state) and the excited state. This restores the energy balance, and also makes the quantum process of de-excitation compatible with the second law of thermodynamics. Indeed, a lower energy state is a state with less entropy as compared to a higher energy state, and the ground state is the least entropy state. This means that the decrease of entropy by de-excitation has to be compensated, and this happens by the sending out of the photon that spreads out in space, and in this way increases the entropy of the compound entity molecule and photon. The entropy reasoning remains valid for the situation that we consider, independent of whether we suppose that the quantum structure in the mind is carried by micro-physical entities or not. This means that a de-excitation, e.g. the concept `pet' that in state $p_1$ decays to the ground state $\hat{p}$, should involve a process of spreading out of a conceptual entity related to `pet'. Our speculation is that speech, apart from the more obvious role it plays in communication between different minds, also fulfills this role. This is probably the reason that if the de-excitation is huge and carries a big emotional energy, speech can function as a catharsis of this emotional energy, which would be why psychotherapy consisting of talking can function quite independent of the content of what is said.

The global and speculative view that can be put forward is the following. The compound of all concepts relevant to a certain individual are stored in memory (a more correct way to say this would be: they are memory) and one specific state of mind of the individual will determine one specific state of this compound of concepts. This state of the compound of concepts is a hugely entangled state, but such that most of the time, the reduced states for each concept apart are the ground states. Any specific context will influence and change the state of mind of the individual, and hence also the entangled state of the compound of concepts, and hence also the ground states of some of the individual concepts. These are the concepts that we will identify as being evoked by this specific context. Most of these changes of state are just excitations that spontaneously will de-excite, such that all the individual concepts are in their ground states again. Form time to time however, a change of state will have consequences that change the structure of the entanglement, or even the structure of some of the concepts themselves. This are the times that the individual learns something new that will be remembered in his or her long term memory, and that will provoke a change of his or her world views. The energetic balance gets redefined when this happens, and a new stable entangled state of the compound of all concepts is introduced, giving rise to new ground states for the individual concepts (for example, `pet' is not any longer what `pet' was before, once  one has his or hers own pet). This new situation, just as the earlier one, is again open to influences of contexts that introduce again the dynamics of excitation and spontaneous de-excitation. 

\section{Summary and Conclusions}
Von Foerster was inclined to push the formalization of whatever happened to interest him at a given time as far as it could go using whatever tools did the job, in order to penetrate into the phenomenon more deeply. In this paper we take a non-operational step, embedding the SCOP in a more constrained structure, the complex Hilbert space, the mathematical space used as a basis of the quantum mechanical formalism. We have good reasons to do so. The generalized quantum formalisms entail the structure of a complete orthocomplemented lattice, and its concrete form, standard quantum mechanics, is formulated within a complex Hilbert space. The SCOP representation of a concept thereby makes strong gains in terms of calculation and prediction power, because it is formulated in terms of the much less abstract numerical space, the complex Hilbert space.

Section \ref{sec:quantummodel} outlines the mathematics of a standard quantum mechanical model in a complex Hilbert space. It is not only the vector
space structure of the Hilbert space that is important, but also the quantum way of using the Hilbert space. A state is described by
a unit vector or a density operator, and a context or property by an orthogonal projection. The quantum formalism furthermore determines the formulas that describe the transition probabilities between states and the weights of the properties. It is by means of these probabilities and weights that we model the typicality values of exemplars and applicability values of properties.

In Section \ref{sec:quantummodel02} we embed the SCOP in a complex Hilbert space, and call the resulting model `the quantum model of a concept', to distinguish it from the more abstract SCOP model. The quantum model is similar to a SCOP model, but it is more precise and powerful because it allows specific numerical predictions. We represented the exemplars, contexts, and states that were tested experimentally for the concept `pet'. Each exemplar is represented as a state of the concept. The contexts, states and properties considered in the experiment are embedded in the complex Hilbert space, where contexts figure as orthogonal projections, states as unit vectors or density operators, and properties as orthogonal projections. The embedding is faithful in the sense that the predictions about frequency values of exemplars and applicability values of properties of the model coincide with the values yielded by the experiment (Section \ref{sec:states/weights}). 

Notice how the so-called `pet fish problem' disappears in our formalism. The pet fish problem refers to the empirical result that a guppy is rated as a good example, not of the concept `pet', nor of the concept `fish', but of the conjunction `pet fish'. This phenomenon, that the typicality of the conjunction is not a simple function of the typicality of its constituent, has come to be known as the `guppy effect', and it cannot be predicted or explained by contemporary theories of concepts. In our experiment, and hence also in the quantum model, we have taken the context `The pet is a fish' to be a context of the concept `pet'. Both the experiment and the quantum model description show the guppy effect appearing in the state of `pet' under the context `The pet is a fish'. Subjects rate guppy as a good example of `pet' under the context `The pet is a fish', and not as a good example of `pet', and the ratings are faithfully described by the quantum model (Section \ref{sec:states/weights}).  Of course this is not the real guppy effect, because we did not yet describe the combination of the concept `pet' and `fish'. Section \ref{sec:conceptualcombination} is devoted to modeling concept combination. 

A specific procedure exists to describe the compound of two quantum entities. The mathematical structure that is used is the structure of the tensor product of the Hilbert spaces that are used to describe the two sub-entities. Section \ref{sec:tensorproduct} outlines the tensor product procedure for quantum entities. The tensor product of Hilbert spaces is a sophisticated structure. One of its curious properties is that it contains elements that are called non-product vectors. The states described in quantum mechanics by these non-product vectors of the tensor product of two Hilbert spaces are the so-called `entangled quantum states'. They describe entanglement between two quantum entities when merging with each other to form a single compound. In the process of working on this quantum representation of concepts, we were amazed to find that it is these very non-products states that describe the most common combinations of concepts, and that more specifically a `pet fish' is described by entangled states of the concepts `pet' and `fish'. This enables us to present a full description of the conceptual combination `pet fish' and hence a solution to the pet fish problem in Section \ref{sec:petfishentangled}. There is more to the tensor product procedure than combining concepts. For example, it allow the modeling of combinations of concepts such as `a pet and a fish', something completely different from `pet fish'. In this case, product states are involved, which means that the combining of concepts by using the word `and' does not entail entanglement (Section \ref{sec:compound}). Finally, we show how our theory makes it possible to describe the combination of an arbitrary number of concepts, and work out the concrete example of the sentence `The cat eats the food' (Section \ref{sec:sentences}).

\bigskip
\noindent
{\bf Acknowledgments:} We would like to thank Alex Riegler and six anonymous reviewers for comments on the manuscript. This research was supported by Grant G.0339.02 of the Flemish Fund for Scientific Research.

\end{document}